\documentclass[11pt,a4paper]{article}
\pdfoutput=1

\usepackage{jcappub}

\usepackage{mathrsfs}
\usepackage{bbm}
\usepackage{bm}
\usepackage{dsfont}
\usepackage{tensor}
\usepackage{xspace}
\usepackage{aeguill}
\usepackage{wasysym}
\usepackage{microtype}
\usepackage{empheq}
\usepackage{ifthen}
\usepackage{amsmath}
\usepackage{subcaption}
\usepackage{wrapfig}
\usepackage{multirow}

\usepackage{graphicx}
\usepackage{dcolumn}
\usepackage{amsfonts,amssymb,slashed,upgreek,color}
\usepackage{multirow}

\newcommand\eps{\ensuremath{\varepsilon}}

\newcommand\define{\equiv}

\newcommand\ex[1]{\mathrm{e}^{#1}}
\renewcommand\i{\ensuremath{\mathrm{i}}}

\newcommand\e[1]{_{\text{#1}}}

\newcommand{\dd}{\mathrm{d}}

\renewcommand\lim[2]{\underset{ #1 \rightarrow #2 }{ \mathrm{lim} } \,}

\newcommand{\delimiters}[4][]{
\ifthenelse{ \equal{#1}{1} }{  #2 #3 #4  }
					{ \ifthenelse{\equal{#1}{2}}{ \big#2 #3 \big#4 }
						{ \ifthenelse{\equal{#1}{3}}{ \Big#2 #3 \Big#4 }
							{ \ifthenelse{\equal{#1}{4}}{ \bigg#2 #3 \bigg#4 }
								{ \ifthenelse{\equal{#1}{5}}{ \Bigg#2 #3 \Bigg#4 }
									{ \left#2 #3 \right#4 }
								}
							}
						}
					}
													}

\newcommand{\pa}[2][]{\delimiters[#1]{(}{#2}{)}}
\newcommand{\pac}[2][]{\delimiters[#1]{[}{#2}{]}}

\newcommand{\HH}{\mathcal{H}}

\newcommand\ee{\end{equation}}
\newcommand\be{\begin{equation}}
\newcommand\eea{\end{eqnarray}}
\newcommand\bea{\begin{eqnarray}}
\newcommand{\bV}{\mathbf{V}}

\newcommand{\bn}{\mathbf{n}}
\newcommand{\bN}{\mathbf{N}}
\newcommand{\bmm}{\mathbf{m}}
\newcommand{\bx}{\mathbf{x}}

\newcommand{\bk}{\mathbf{k}}
\newcommand{\hbk}{\hat{\mathbf{k}}}
\newcommand{\mF}{\mathcal{F}}

\newcommand{\B}{\textrm{B}}
\newcommand{\F}{\textrm{F}}
\renewcommand{\L}{\textrm{L}}
\newcommand{\M}{\textrm{M}}
\newcommand{\N}{\textrm{N}}
\renewcommand{\P}{\textrm{P}}
\newcommand{\BF}{\textrm{BF}}

\newcommand{\bB}{b_\textrm{B}}
\newcommand{\bF}{b_\textrm{F}}
\newcommand{\fe}{f^\textrm{evol}}
\newcommand{\hNe}{\hat{\mathcal{N}}_\textrm{E}}
\newcommand{\hMe}{\hat{\mathcal{M}}_\textrm{E}}
\newcommand{\Ne}{\mathcal{N}_\textrm{E}}
\newcommand{\Me}{\mathcal{M}_\textrm{E}}
\newcommand{\hxiq}{\hat\xi_2}
\newcommand{\hxih}{\hat\xi_4}
\newcommand{\hxid}{\hat\xi_1^{\rm BF}}

\newcommand{\xid}{\xi_1^{\rm BF}}
\newcommand{\Pstr}{P^{\rm str}}
\newcommand{\VV}{\mathcal{V}}
\newcommand{\Eb}{E^{\rm break}}

\newcolumntype{C}[1]{>{\centering\arraybackslash}p{#1}}
\newcolumntype{L}[1]{>{\raggedright\arraybackslash}p{#1}}
\newcolumntype{R}[1]{>{\raggedleft\arraybackslash}p{#1}}

\newlength{\boxtitlelength}
\newlength{\halfrulelength}
\newcommand{\boxtitle}[1]{\footnotesize\bf{\:#1\:}}

\definecolor{blue4}{RGB}{0,0,143}
\definecolor{red4}{RGB}{143,0,0}
\definecolor{orange}{RGB}{255,128,0}
\definecolor{darkcyan}{RGB}{0,128,128}
\definecolor{olive}{RGB}{0,128,0}
\definecolor{purple}{RGB}{128,0,128}
\definecolor{cyan2}{RGB}{0,255,255}
\definecolor{fushia}{RGB}{255,0,255}
\definecolor{mygray}{gray}{0.5}
\definecolor{lightgray}{gray}{0.85}

\makeatletter
\def\@fpheader{\relax}
\makeatother

\title{A null test of the equivalence principle using relativistic effects in galaxy surveys}

\author[a]{Camille Bonvin,}
\author[a]{Felipe Oliveira Franco,}
\author[a,b]{Pierre Fleury}

\affiliation[a]{D\'{e}partment de Physique Th\'{e}orique and Center for Astroparticle Physics, Universit\'{e} de Gen\`{e}ve,\\
24 quai Ernest-Ansermet, 1211 Gen\`{e}ve 4, Switzerland}

\affiliation[b]{Instituto de F\'isica Te\'orica UAM-CSIC,
Universidad Auton\'oma de Madrid,\\
Cantoblanco, 28049 Madrid, Spain}

\emailAdd{camille.bonvin@unige.ch}
\emailAdd{felipe.oliveira@unige.ch}
\emailAdd{pierre.fleury@uam.es}

\abstract{The weak equivalence principle is one of the cornerstone of general relativity. Its validity has been tested with impressive precision in the Solar System, with experiments involving baryonic matter and light. However, on cosmological scales and when dark matter is concerned, the validity of this principle is still unknown. In this paper we construct a null test that probes the validity of the equivalence principle for dark matter. Our test has the strong advantage that it can be applied on data without relying on any modelling of the theory of gravity. It involves a combination of redshift-space distortions and relativistic effects in the galaxy number-count fluctuation, that vanishes if and only if the equivalence principle holds. We show that the null test is very insensitive to typical uncertainties in other cosmological parameters, including the magnification bias parameter, and to non-linear effects, making this a robust null test for modified gravity.}


\date{\today}

\begin{document}

\maketitle
\flushbottom

\section{Introduction}
\label{sec:introduction}

The distribution of galaxies in redshift space is a highly sensitive probe of the theory of gravity, that can be used to look for deviations from general relativity (GR). The usual way to probe such deviations is to measure the two-point correlation function of galaxies (or its Fourier transform, the power spectrum), and confront these measurements with a theoretical modelling that accounts for modifications of gravity. This can be done in two different ways. The simplest possibility is to calculate the correlation function in a specific model of dark energy or modified gravity, like for example $f(R)$ gravity~\citep{1970MNRAS.150....1B,1980PhLB...91...99S}, and use observations to place constraints on the parameters of the model.  The second approach consists in parameterizing deviations from GR directly at the level of the correlation function. One well-known example is the $\gamma$ parameterization of the growth rate, which is directly measured through the multipoles of the correlation function:  $f(z)=\Omega_{\rm m}(z)^\gamma$, where $\gamma$ is a free parameter that takes the value $\gamma\simeq0.55$ in GR~\citep{Wang_1998,PhysRevD.72.043529}. 

Recently, various frameworks have been developed to combine these two approaches, like the Effective Theory of Dark Energy~\citep{Gubitosi:2012hu} and the Parameterized Post-Friedmann approach~\citep{Baker:2012zs}. These frameworks provide general parameterizations of deviations from GR describing a large class of theories, and whose parameters directly affect large-scale structure observables. They can therefore be used to consistently search for deviations from GR. These parameterizations have nevertheless two drawbacks. First, in order to be as general as possible, they feature various free functions of time. This results in too many degrees of freedom that cannot all be constrained by observations. Second, even if these parameterizations are very general, they do not encompass all possible deviations from GR. Hence, by using them, we automatically restrict to some specific class of theories. 

In this paper, we envisage an even more agnostic approach, which consists in testing specific properties of gravity, rather than a particular set of alternative models. Namely, we propose \emph{a null test to probe the weak equivalence principle} for dark matter. The weak equivalence principle, which states that all objects fall in the same way in a gravitational potential, is one of the cornerstone of GR. This principle is extremely well tested in the Solar System, with experiments involving baryonic matter and light. The validity of this principle is however much more difficult to probe on cosmological scales, or when the unknown dark matter is concerned.

To be specific, we will test whether the motion of dark matter on cosmic scales is governed by \emph{Euler's equation}. This is indeed a test of the equivalence principle in a generalised sense. Let us elaborate on this point. Suppose that there exists a description of gravity such that electromagnetism\footnote{We focus here on electromagnetism, since as we will see below, we use photons to measure the gravitational potential which enters into Euler's equation.} 
is mimimally coupled to the geometry of space-time. While this is always true in GR, it may require some re-parameterisation in alternative theories; for example, in scalar-tensor theories it would correspond to working in the Jordan frame of electromagnetism. In that framework, light travels along null geodesics. \emph{If} dark matter is coupled to gravity just like electromagnetism is, i.e.\ if the equivalence between these species holds, \emph{then} dark matter must travel along time-like geodesics, and hence its cosmic motion is dictated by Euler's equation. This remains true even in the presence of a fifth force. In a theory of gravity where both electromagnetism and dark matter are non-minimally coupled to a new gravitational degree of freedom (e.g.\ a scalar field), they are both affected by this new coupling, which propagates an additional force. This fifth force does however not violate Euler's equation, since it is always possible to rewrite the theory in the Jordan frame where light and dark matter follow geodesics of the metric. Hence a violation of Euler's equation implies necessarily a \emph {difference} between the way photons and dark matter fall, i.e.\ a violation of the equivalence principle.

The phenomenology that may lead to such a violation is very rich and diverse. Let us mention a few examples. Dark matter could be non-minimally coupled to space-time geometry~\cite{Gleyzes:2015pma}, while light would not. In this case, in the Jordan frame of electromagnetism, dark matter would not follow time-like geodesiscs of the metric. Such a non-universal coupling could also emerge from screened modifications of GR~\cite{Hui:2009kc}, from violations of Lorentz invariance (existence of a preferred frame) which could affect relativistic and non-relativistic particles differently, or from violations of the strong equivalence principle (Nordtvedt effect~\cite{PhysRev.169.1014}). The latter option would be particularly relevant if a large fraction of dark matter were made of compact objects, such as primordial black holes~\cite{Clesse:2017bsw}. Finally, departures from geodesic motion could be due to departures from free-fall on cosmic scales. Interesting scenarios would then include exotic dark matter-dark energy interactions, either from the sole exchange of energy~\cite{Wands:2012vg}, or via momentum transfer~\cite{Asghari:2019qld}. In that sense, testing Euler's equation for dark matter relates to a wide range of new physics.

The standard analysis of redshift-space distortions usually applies the equivalence principle in order to translate the observed galaxies' peculiar velocities into measurements of the time component of the metric, $\Psi$. However, if dark matter does not obey the equivalence principle, the peculiar velocities of galaxies, which are governed by the peculiar velocities of the dark matter halos, will not satisfy Euler's equation. The relation between peculiar velocities and the gravitational potential $\Psi$ is consequently altered. In~\cite{Bonvin:2018ckp}, we proposed a general parameterization of deviations from Euler's equation, that would be due to a violation of the equivalence principle. We then showed that such a violation can be tested using the dipole of the cross-correlation function between bright and faint galaxies. The dipole provides indeed a direct measurement of the gravitational potential $\Psi$, through the effect of gravitational redshift~\cite{Bonvin:2013ogt}. Combining a measurement of the dipole with that of the monopole, quadrupole and hexadecapole of the correlation function consequently allows one to constrain deviations from Euler's equation. 

In this paper, we extend our previous analysis by constructing a null test of the equivalence principle. We combine observables in such a way that the null test vanishes if and only if Euler's equation is valid. Such a test has the advantage that it does not require any specific modelling of the theory of gravity. In particular, it can be applied to the data without assuming a parameterization of Euler's equation. One simply combines observables and concludes whether or not the equivalence principle holds, based on the outcome of the test. Therefore, our test provides a robust way of testing one of the most fundamental properties of GR. Note that the null test specifically probes the equivalence of free fall between dark matter and light. It relies indeed on comparing the peculiar velocity of galaxies (which is governed by the peculiar velocity of dark matter haloes) and the gravitational potential $\Psi$ which is measured via gravitational redshift, i.e.\ which probes the way photons escape the gravitational potential.
As such, this test is complementary to other tests of the equivalence principle proposed in~\cite{Kehagias:2013rpa,Creminelli:2013nua}, based on consistency relations between the two-point and three-point correlation functions, that test the equivalence of free fall between dark matter and baryons.

The rest of the paper is organised as follows. In Sec.~\ref{sec:nulltest} we define our large-scale structure observables and construct the null test. In Sec.~\ref{sec:cov}, we calculate the variance of the null test, which is essential to assess its sensitivity to a violation of the equivalence principle. In Sec.~\ref{sec:results}, we forecast the performance of the null test for a survey like SKA phase 2. In Sec.~\ref{sec:contaminations}, we study the possible contaminations to the null test, and we conclude in Sec.~\ref{sec:conclusion}.

\section{The null test}
\label{sec:nulltest}

We build the null test from the galaxy number-count fluctuation $\Delta(z,\bn)$, observed at redshift $z$ and in direction $\bn$. The dominant contributions to $\Delta$ are due to density perturbations and to redshift-space distortions (RSD)
\be
\Delta^{\rm st}= b \delta - \frac{1}{\HH} \, \partial_r(\bV\cdot\bn)\, ,\label{Deltast}
\ee
where $b$ is the bias, $\HH$ is the Hubble factor in conformal time $\eta$ and $r$ is the comoving radial coordinate in direction $\bn$. $\delta$ denotes the matter density perturbation, and $\bV$ is the galaxy peculiar velocity. Since a galaxy always sits inside a dark matter halo, the halo exerts a binding force on the galaxy, and consequently we expect the velocity of the galaxy to be given by the velocity of the dark matter halo, even if dark matter and baryonic matter experience gravitation in a different way, due to the breaking of the equivalence principle. As a consequence we assume that $\bV_{\rm dm}=\bV$.

In addition to these standard contributions, $\Delta$ is affected by various relativistic effects. Among those, the dominant contributions are given by~\cite{Yoo:2009au,Bonvin:2011bg,Challinor:2011bk}
\be
\Delta^{\rm rel}= \left(1-5s+\frac{5s-2}{r\HH}-\frac{\HH'}{\HH^2}+\fe \right) \bV\cdot\bn+\frac{1}{\HH}\bV'\cdot\bn+\frac{1}{\HH}\partial_r\Psi\, , \label{Deltarel}
\ee
where the first two terms are Doppler contributions and the last term is the contribution from gravitational redshift. $\Psi$ is the gravitational potential involved in the time component of the metric\footnote{We work with the metric $\dd s^2=a^2(\eta)\big[-(1+2\Psi)\dd\eta^2+(1-2\Phi)\dd\bx^2 \big]$.}, a prime denotes a derivative with respect to conformal time $\eta$, $s$ is the magnification bias parameter, related to the slope of the galaxy luminosity function, and $\fe$ is the evolution bias. $\Delta^{\rm rel}$ is suppressed by one power $\HH/k$ compared to the dominant standard contributions $\Delta^{\rm st}$. In addition to the terms in Eq.~\eqref{Deltarel}, $\Delta$ contains other relativistic effects that are suppressed by $(\HH/k)^2$ with respect to $\Delta^{\rm st}$, and that are therefore negligible in the regime we are interested in here. 

\begin{figure}[!t]
\centering
\includegraphics[width=0.65\columnwidth]{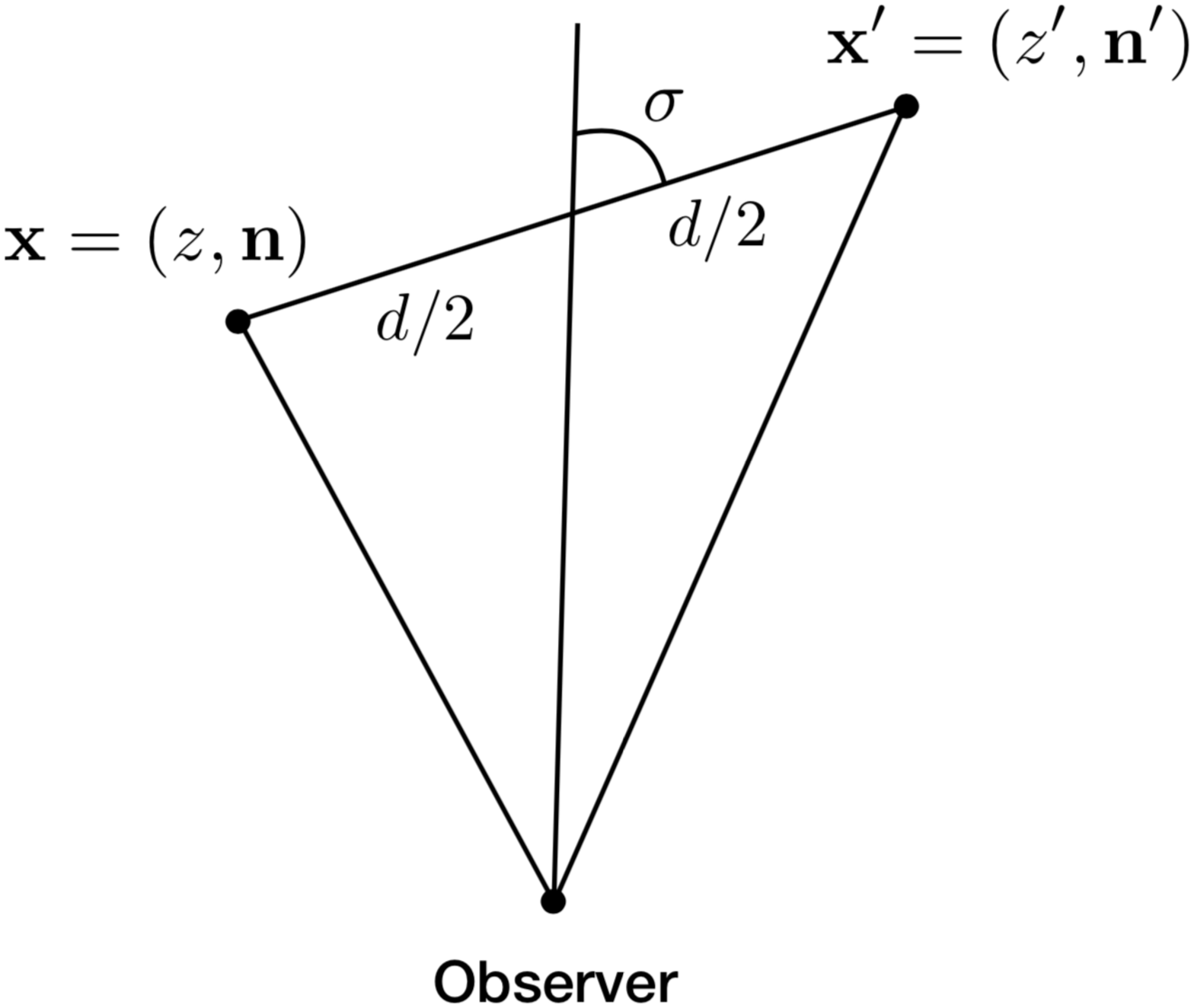}
\caption{Definition of the angle $\sigma$, used to extract the multipoles of the correlation function.}
\label{fig:triangle}
\end{figure}

If dark matter obeys the equivalence principle, then the galaxy peculiar velocity is related to the gravitational potential $\Psi$ by Euler's equation
\be
\bV'+\HH\bV+\nabla\Psi=0\, .\label{Euler}
\ee
Our goal is to find a combination of $\Delta^{\rm st}$ and $\Delta^{\rm rel}$ that vanishes if Euler's equation holds. More precisely, since only the radial part of $\bV$ contributes to $\Delta$, and since one always measures correlations of $\Delta$, we are looking for a combination of observables that is proportional to
\begin{align}
& \big\langle(\bn\cdot\bV')\cdot\Delta^{\rm st}  \big\rangle(d,\mu, z)+
\HH\big\langle (\bn\cdot\bV)\cdot\Delta^{\rm st} \big\rangle(d,\mu, z)
+\big\langle \partial_r\Psi\cdot\Delta^{\rm st}\big\rangle(d,\mu, z)\, ,
\label{comb}
\end{align}
where $d$ denotes the separation between the two pixels that are correlated, $\mu$ is the cosine of the angle between the pair of pixels and the direction of observation, and $z$ is the mean redshift of the pair, see~Fig.~\ref{fig:triangle}. From Eqs.~\eqref{Deltast} and~\eqref{Deltarel} we see that such a combination can be found if we can measure separately $\langle \Delta^{\rm st}\Delta^{\rm st}\rangle$ and $\langle \Delta^{\rm st}\Delta^{\rm rel}\rangle$.
This can be done by extracting the multipoles of $\langle\Delta \Delta\rangle$, i.e.\ averaging $\langle\Delta\Delta\rangle(d,\mu, z)$ over the orientation of the pair $\mu$, weighted by the appropriate Legendre polynomial of $\mu$.

The monopole, quadrupole and hexadecapole of the correlation function, $\langle\Delta \Delta\rangle$, are mainly affected by $\langle \Delta^{\rm st}\Delta^{\rm st}\rangle$. The relativistic effects contribute only through quadratic terms $\langle\Delta^{\rm rel}\Delta^{\rm rel} \rangle$ by symmetry. As a consequence, their impact on the even multipoles is suppressed by $(\HH/k)^2$ and can be safely neglected. The monopole of $\langle\Delta \Delta\rangle$ is sensitive to the isotropic part $\delta\delta$, which does not enter into Eq.~\eqref{comb}; it is thereby not appropriate to construct the null test. The quadrupole and hexadecapole on the other hand are sensitive to density-velocity correlations and velocity-velocity correlations, that both enter into Eq.~\eqref{comb}.

The correlation $\langle \Delta^{\rm st}\Delta^{\rm rel}\rangle$ does not contribute to the even multipoles of $\langle\Delta \Delta\rangle$ by symmetry. However it does contribute to the odd multipoles. These odd multipoles exist only if we cross-correlate different populations of galaxies, like for example a bright population, B, (whose luminosity is higher than a given threshold) and a faint population, F. In this case, the dipole and octupole of $\langle\Delta_\B\Delta_\F\rangle$ are proportional to $\langle \Delta_\B^{\rm st}\Delta_\F^{\rm rel}\rangle+\langle \Delta_\F^{\rm st}\Delta_\B^{\rm rel}\rangle$, as shown in~\cite{Bonvin:2013ogt}. Hence, by combining even and odd multipoles, and using two populations of galaxies, we can construct a combination which is proportional to Eq.~\eqref{comb}, i.e.\ a combination that exactly vanishes if dark matter obeys the equivalence principle.

We start by splitting the population of galaxies into two populations with different luminosities. The bright population contains all galaxies with a luminosity higher than a given threshold and has mean bias $\bB$, magnification bias $s_\B$ and evolution bias $\fe_\B$. The faint population contains all fainter galaxies, and has bias $\bF$, magnification bias $s_\F$ and evolution bias $\fe_\F$. Given this split, we can then measure separately the multipoles of the bright population, of the faint population, and the multipoles of the cross-correlation between bright and faint. 

In the flat-sky approximation, the three different quadrupoles are given by
\begin{align}
\xi^{\B\B}_2 &=  \pac{ -\frac{4}{3} \bB f - \frac{4f^2}{7} } \mu_2(d,z)\, , \label{quadBB}\\
\xi^{\F\F}_2 &=  \pac{ -\frac{4}{3}\bF f - \frac{4f^2}{7} } \mu_2(d,z)\, , \label{quadFF}\\
\xi^{\B\F}_2 &=  \pac{ -\frac{2}{3} \pa{b\e{B}+b\e{F}}f - \frac{4f^2}{7} } \mu_2(d,z) \label{quadBF}\, .
\end{align}
The hexadecapole being independent of the bias, it is the same for all correlation functions,
\be
\xi^{\B\B}_4=\xi^{\F\F}_4=\xi^{\B\F}_4=\xi_4=\frac{8f^2}{35} \, \mu_4(d,z)\, .\label{hexa}
\ee
The functions $\mu_2$ and $\mu_4$ are given by
\be
\mu_\ell(d,z)=\frac{1}{2\pi^2}\int \dd k \; k^2 P(k,z)j_\ell(kd)\, ,\label{muell}
\ee 
where $P(k,z)$ is the density power spectrum.
To derive Eqs.~\eqref{quadBB} to~\eqref{hexa}, we have made two assumptions. First we have assumed that the continuity equation for matter is valid. This means that we only consider theories in which there is no flow of energy from dark matter to another component.\footnote{In particular, interacting CDM-vacuum scenarios~\cite{Wands:2012vg, Hogg:2020rdp} are excluded.} In this case, the velocity potential in Fourier space is related to the density at sub-horizon scales by
\be
V(\bk, z)=-\frac{\HH(z)}{k} f(z)\delta( \bk,z)\, ,
\ee
where the growth rate $f$ is defined as
\be
f(z)=\frac{\dd\ln D_1(z)}{\dd\ln a}\,,
\ee
with $D_1$ the linear growth. The second assumption needed to obtain Eqs.~\eqref{quadBB} to~\eqref{hexa} is that $D_1$ and $f$ are scale-independent, i.e.\ that they do not depend on $k$. We will discuss in the next section what happens if the growth rate depends on scale.

The dipole and octupole exist only when we cross-correlate the bright and faint population. Since they are proportional to $\langle \Delta_\B^{\rm st}\Delta_\F^{\rm rel}\rangle+\langle \Delta_\F^{\rm st}\Delta_\B^{\rm rel}\rangle$, they are suppressed by one power of $\HH/k$ with respect to the even multipoles. This makes them more difficult to measure than the even multipoles. However, as shown in~\cite{Bonvin:2015kuc}, the odd multipoles will be detectable with the coming generation of galaxy surveys. To construct the null test, we only consider the dipole, which has a significantly higher signal-to-noise ratio than the octupole (though see~\cite{Kodwani:2019onz} for a discussion about the use of the octupole to test the equivalence principle). 

The dipole can be extracted from the cross-correlation of bright and faint galaxies in the following way
\be
\xid=\frac{3}{2}\int_{-1}^1\dd\mu\; P_1(\mu)\frac{1}{2}\Big[\langle\Delta_\B(z,\bn)\Delta_\F(z',\bn')\rangle- \langle\Delta_\F(z,\bn)\Delta_\B(z',\bn')\rangle\Big]\, , \label{dipdef}
\ee
where $P_1(\mu)=\mu$ is the first-order Legendre polynomial, and $\mu=\cos\sigma$, $\sigma$ being the angle formed by the median of the triangle (O,$\bx$,$\bx'$) emerging from O, and the axis connecting $\bx=(z,\bn)$ to $\bx'=(z,\bn')$, as depicted in Fig.~\ref{fig:triangle}. The two terms in the integrand of Eq.~\eqref{dipdef} translate the fact that we need to consider both the bright and the faint population in each pixel; the minus sign ensures that we target relativistic effects, which are anti-symmetric under the exchange of $\bx$ and $\bx'$. 

The relativistic contributions to $\Delta$ can be rewritten as
\be
\Delta_\L^{\rm rel}=\frac{1}{\HH}\Big[\bV'\cdot\bn+\HH \bV\cdot\bn+ \partial_r\Psi\Big]+ \left(-5s_\L+\frac{5s_\L-2}{r\HH}-\frac{\HH'}{\HH^2}+\fe_\L \right) \bV\cdot\bn\, , 
\ee
where $\L=\B,\F$. The term in square bracket vanishes if Euler's equation~\eqref{Euler} is valid. To allow for a breaking of the equivalence principle for dark matter, we define 
\be
\Eb(z,\bn)=\bV'\cdot\bn+\HH \bV\cdot\bn+ \partial_r\Psi\, ,
\ee
so that the relativistic contributions become
\be
\Delta_\L^{\rm rel}=\frac{1}{\HH}\Eb+ \left(-5s_\L+\frac{5s_\L-2}{r\HH}-\frac{\HH'}{\HH^2}+\fe_\L \right) \bV\cdot\bn\, . \label{Deltarelbreak}
\ee
$\Eb$ is by construction the product of a vector field (which modifies Eq.~\eqref{Euler}), with the direction of observation $\bn$. As such, the correlation of $\Eb$ with $\Delta^{\rm st}$ is necessarily anti-symmetric under the exchange of $\bx$ and $\bx'$; thus, it contributes to the dipole.
 
Inserting~\eqref{Deltarelbreak} into~\eqref{dipdef}, we obtain 
\begin{align}
\xi^{\B\F}_1 =& \frac{\HH}{\HH_0} 
\Bigg[(b_\B-b_\F) f \pa{ \frac{2}{r\HH} + \frac{\HH'}{\HH^2}}
+ 3 (s_\F-s_\B) f^2 \pa{ 1-\frac{1}{r\HH} }+ 5 (b_\B s_\F-b_\F s_\B) f \pa{ 1-\frac{1}{r\HH}}\nonumber\\
&+\frac{3}{5}\left(\fe_\B-\fe_\F\right)f^2
+\left(b_\F\fe_B-b_\B\fe_F\right)f\Bigg] \nu_1(d,z)-\frac{2}{5} (b\e{B}-b\e{F}) f \frac{d}{r}  \mu_2(d,z) \nonumber\\
&+\frac{3(b_\B-b_\F)}{4\HH}\int_{-1}^1\dd\mu \; P_1(\mu)\langle \delta\cdot \Eb \rangle(d,\mu,z)\, ,\label{xirel}
\end{align}
where
\be
\nu_1(d,z)=\frac{1}{2\pi^2}\int \dd k \; k\HH_0 P(k,z)j_1(kd)\, . \label{nu1}
\ee
The last line in Eq.~\eqref{xirel} clearly vanishes if Euler equation is valid. The last term in the second line of Eq.~\eqref{xirel} is the wide-angle contribution from the standard terms, which contaminates the relativistic dipole and hence needs to be included. Note that this wide-angle contribution is the main reason why we are using the correlation function instead of the power spectrum. In the correlation function, wide-angle effects can be consistently computed and included in the theoretical modelling. On the other hand, in the power spectrum, the flat-sky approximation is used from the beginning, and wide-angle effects are by construction exactly zero (except if one uses different line-of-sights for each pair as proposed in~\cite{Yamamoto:2005dz}, but this is not straightforward). Since the contamination from wide-angle effects is quantitatively comparable to the relativistic effects, it would be inconsistent to neglect them.

Let us now proceed and derive a combination of the quadrupoles, hexadecapole, and dipole, that is proportional to the last line of Eq.~\eqref{xirel}. 
There are, actually, several possibilities. The simplest one depends on the auto-correlation quadrupoles, $\xi_2^{\B\B}$ and $\xi_2^{\F\F}$, but not on the cross-correlation quadrupole $\xi_2^{\B\F}$:
\begin{align}
\hNe&= A\,\hxid+C_\B\,\hxiq^{\B\B}+C_\F\,\hxiq^{\F\F}+D\,\hxih\, , \label{Ne}
\end{align}
with the four coefficients
\begin{align}
A&= \frac{\bar \HH_0}{\bar \HH}\frac{1}{\bar\nu_1(d,z)}\, ,\label{coefftest}\\
C_\B&=\left[\frac{3}{4}\left(\frac{2}{\bar r\bar\HH}+\frac{\bar\HH'}{\bar\HH} \right)+
\frac{15}{4}\left(1-\frac{1}{\bar r\bar\HH}\right)s_\F-\frac{3}{4}\fe_\F \right]\frac{1}{\bar\mu_2(d,z)}
-\frac{3}{10}\frac{\bar{\HH}_0}{\bar{\HH}}
\frac{d}{\bar{r}}\frac{1}{\bar\nu_1(d,z)}
\, ,\nonumber\\
C_\F&=-\left[\frac{3}{4}\left(\frac{2}{\bar r\bar\HH}+\frac{\bar\HH'}{\bar\HH} \right)+
\frac{15}{4}\left(1-\frac{1}{\bar r\bar\HH}\right)s_\B -\frac{3}{4}\fe_\B\right]\frac{1}{\bar\mu_2(d,z)}
+\frac{3}{10}\frac{\bar{\HH}_0}{\bar{\HH}}
\frac{d}{\bar{r}}\frac{1}{\bar\nu_1(d,z)}
\, ,\nonumber\\
D&=\left[\frac{15}{4}\left(1-\frac{1}{\bar r\bar\HH}\right)(s_\B-s_\F)+\frac{3}{4}\left(\fe_\F-\fe_\B \right)\right]\frac{1}{\bar\mu_4(d,z)}\, .\nonumber
\end{align}
In the above expressions, a bar indicates quantities that are calculated in a fiducial $\Lambda$CDM cosmology. The null test, $\hNe$, is thereby a combinations of observables (the multipoles), and theoretical coefficients calculated in a $\Lambda$CDM cosmology. If the fiducial cosmology is a correct description of the actual Universe, then the ensemble average of $\hNe$ reduces to
\begin{align}
\Ne=\langle\hNe\rangle=\frac{\HH_0}{\HH^2}\frac{3(b_\B-b_\F)}{4\bar\nu_1(d,z)}\int_{-1}^1\dd\mu \; P_1(\mu)\langle \delta\cdot \Eb \rangle(d,\mu,z)\, .\label{meanN}
\end{align}
Therefore, $\Ne$ vanishes if $\Eb=0$, i.e.\ if the equivalence principle holds, and it differs from zero if the equivalence principle is violated. This statement is completely independent on the mechanism that would be responsible for such a violation. It does not rely on any modelling of $\Eb$, and can be applied to the data without any assumption on the theory of gravity.

The combination $\hNe$ is not the only one which is proportional to $\Eb$. Since the quadrupoles are related by
\begin{equation}
\hxiq^{\B\B} + \hxiq^{\F\F} = 2 \hxiq^{\B\F} \, ,
\end{equation}
any combination of the form
\be
\hNe
+\frac{\lambda(d,z)}{\bar{\mu}_2(d,z)}
\times
\pa{
\hxiq^{\B\B}
+ \hxiq^{\F\F}
- 2 \hxiq^{\B\F}
}\,,
\label{lambda}
\ee
with $\lambda$ an arbitrary function of $d$ and $z$,
is also proportional to $\Eb$. Ideally, one would like to find the $\lambda$ that optimises the sensitivity of the null test to a violation of the equivalence principle. This turns out to be very challenging in practice, because optimisation requires to express the variance of the null test as a function of $\lambda$, and then to maximise the Fisher matrix which depends on the the inverse of that covariance matrix. Such a procedure is beyond the scope of the present paper.

If we set $\lambda$ to be the following function of $z$ only, the resulting alternative null test features the same coefficient in front of both $\hxiq^{\B\B}$ and $\hxiq^{\F\F}$,
\begin{equation}
\lambda(z)
= \frac{15}{8} \pa{1-\frac{1}{\bar{r}\bar{\HH}}}
    (s_\B-s_\F)+\frac{3}{8}\left(\fe_\F-\fe_\B\right)\, .
\end{equation}
The alternative test reads
\begin{align}
\hMe=A\,\hxid+C\,\hxiq^{\B\B}-C\,\hxiq^{\F\F}+C_{\B\F}\,\hxiq^{\BF}+D\,\hxih\, ,\label{Me}
\end{align}
where $A$ and $D$ are the same as before, and
\begin{align}
C=&\left[\frac{3}{4}\left(\frac{2}{\bar r\bar\HH}+\frac{\bar\HH'}{\bar\HH} \right)+
\frac{15}{8}\left(1-\frac{1}{\bar r\bar\HH}\right)(s_\B+s_\F)-\frac{3}{8}\left(\fe_\B+\fe_\F\right) \right]\frac{1}{\bar\mu_2(d,z)}\label{coefftest2}\\
&-\frac{3}{10}\frac{\bar{\HH}_0}{\bar{\HH}}
\frac{d}{\bar{r}}\frac{1}{\bar\nu_1(d,z)}\,,\nonumber \\
C_{\B\F}&=\left[\frac{15}{4}\left(1-\frac{1}{\bar r\bar\HH}\right)(s_\F-s_\B)+\frac{3}{4}\left(\fe_\B-\fe_\F\right)\right]\frac{1}{\bar\mu_2(d,z)}\, .\nonumber
\end{align}
The ensemble average of $\hMe$ is the same as that of $\hNe$, given by the right-hand side of Eq.~\eqref{meanN}.

Let us now explore under which conditions our null tests $\Ne$ and $\Me$ may be non-vanishing. Different scenarios can be identified. Two of them actually teach us about gravity, but the others are spurious, i.e.\ due to some contamination that is independent on the physics that we aim to probe.

\paragraph{Violation of the equivalence principle.} The first scenario is obviously $\Eb\neq 0$, i.e.\ if the equivalence principle is violated. This is the very reason why the test was designed.

\paragraph{Velocity bias.} The second scenario where $\Ne, \Me\not=0$ occurs is in the case where the velocity of galaxies is biased. Different effects can generate a velocity bias. The first one is related to the population of galaxies that we observe. If those galaxies are satellite galaxies, we expect them to be moving with respect to the dark matter halos~\cite{Berlind:2002rn}. As such their velocity does not trace the dark matter velocity, and they do not obey Euler's equation. This bias is however relevant only at small separations~\cite{Reid:2014iaa} and it does not affect the linear scales considered in this work. The second effect that would generate a velocity bias is if dark matter halos do not move in the same way as dark matter particles, due for example to the presence of friction terms. Such a bias has been measured in numerical simulations~\cite{Colin:1999bh}, and affects only small scales. It is therefore also irrelevant for our null test. Finally, a last possibility is a velocity bias coming from the fact that we observe only peaks of the density field (where galaxies are formed), and that consequently the observed velocity field is not a fair representation of the matter flow~\cite{Biagetti:2015hva}. Such a statistical bias would however not modify the relation between the velocity of one particular galaxy and the gravitational potential associated with this motion. It would only modify the relation between the averaged velocity of galaxies and the averaged gravitational potential of the sample. As such it would have no effect on the null test, which is sensitive to the validity of Euler's equation for pairs of galaxies sitting in the same gravitational potential.

\paragraph{Wrong power spectrum.} The third scenario leading to non-zero $\Ne, \Me$ is if the $\Lambda$CDM power spectrum used to calculate $\bar\mu_2, \bar\mu_4$ and $\bar\nu_1$ in Eqs.~\eqref{coefftest} and~\eqref{coefftest2} differs from the true one. This can happen for different reasons. The first one is if the power spectrum at early time (for example at recombination) differs from the fiducial one. This would mean that the true values of the cosmological parameters $\Omega\e{b},\Omega\e{m}$, $h$ and $n\e{s}$ differ from the fiducial ones. These parameters are however strongly constrained by Planck, and we sill show in Sec.~\ref{sec:contaminations} that if their true value is consistent with the best-fit from Planck at 3$\sigma$, then the deviations from zero generated in $\Ne$ and $\Me$ remain significantly smaller than their variance and do not invalidate the null tests. 

The second reason for the true power spectrum to differ from the fiducial one is if the evolution of structures at late time (once the acceleration of the Universe has started) differs from the one in $\Lambda$CDM. Since the null tests are targeted at probing models beyond $\Lambda$CDM, we need to understand which features will generate a non-zero $\Ne$ and $\Me$ besides a violation of the equivalence principle. Clearly, if the true power spectrum is related to the fidudical one by a scale-independent factor
\be
\label{eq:Pevol}
P(k, z)=\left[\frac{D_1(z)}{\bar D_1(z)}\right]^2\bar P(k, z)\, ,
\ee
then $\left[{D_1(z)}/\bar D_1(z)\right]^2$ can be factorised out of Eqs.~\eqref{Ne} and \eqref{Me}, so that $\Ne=\Me=0$. So, any dark energy or modified gravity model that preserves the equivalence principle \emph{and} the scale-independence of the growth of structure, produces a zero null test, even if $D_1$ is different from the one in $\Lambda$CDM. 

On the other hand, if the growth of structure depends on scale, $D_1(k,z)$ cannot be taken out of the integrals in Eqs.~\eqref{muell} and~\eqref{nu1} and $\Ne$ and $\Me$ generally become non-zero. Therefore, $\Ne$ and $\Me$ are not only a test of the equivalence principle, but also a test of the scale-independence of the growth of structure. However, we will see in Sec.~\ref{sec:contaminations} that $\Ne, \Me$ are not the best combinations to test the scale-independence of the growth of structure. As shown in~\cite{Franco:2019wbj}, there exists another combination, $\mathcal{N}_f$, which is more sensitive to a breaking of scale-independence. This means that any observed deviation from zero with $\Ne$ and $\Me$ would lead to a larger observed deviation in $\mathcal{N}_f$. Moreover, $\mathcal{N}_f$ is not sensitive at all to a breaking of the equivalence principle, since it does not involve the gravitational potential $\Psi$. As a consequence, if we see no deviations in $\mathcal{N}_f$, but we observe a non-zero $\Ne$ and $\Me$, this necessarily means that the equivalence principle is broken. 

\paragraph{Wrong background parameters.} Finally, the last reason for $\Ne$ and $\Me$ not to vanish is if the fiducial $\Lambda$CDM cosmology used to calculate the background quantities $\bar r$ and $\bar\HH$ in Eq.~\eqref{coefftest} differs significantly from the true background cosmology. In Section~\ref{sec:contaminations} we will see that if the background is consistent with Planck constraints at 3$\sigma$, then the deviations from zero in $\Ne$ and $\Me$ are much smaller than their variance, which means that they do not invalidate the null test. In addition, the coefficients~\eqref{coefftest} also depend on the slope of the galaxy luminosity function $s_\B$ and $s_\F$. These coefficients can be measured for each of the populations. In Sec.~\ref{sec:contaminations} we will see that an uncertainty of $10\,\%$ or less in the measurement of $s_\B$ and $s_\F$ generates deviations in $\Ne$ and $\Me$ that are smaller than their variance. Similarly, the coefficients depend on the evolution biases $\fe_\B$ and $\fe_\F$. The value of these biases is uncertain, but they can in principle be measured by looking at the evolution of the number densities of galaxies with redshift. In Sec.~\ref{sec:contaminations} we will see that an uncertainty of 40\% or less in the measurement of $\fe_\B$ and $\fe_\F$ generates deviations in $\Ne$ and $\Me$ that are smaller than the variance, so that the null test is not invalidated.

\medskip

To summarise, the only effects that can lead the ensemble average of $\hNe, \hMe$ to be larger than their variance, i.e.\ to a statistically significant rejection of the null hypothesis, is if the growth of structure differs from $\Lambda$CDM in a scale-dependent way, or if the equivalence principle is violated. To differentiate between these two cases, one may perform the additional test $\mathcal{N}_f$ defined in~\cite{Franco:2019wbj}, which is only sensitive to a scale-dependent growth. Note that Horndeski models do produce a scale-independent growth of structures in the quasi-static regime, but they can break the equivalence principle if dark matter and baryonic matter are coupled differently to the scalar field.

\section{Variance}
\label{sec:cov}

The sensitivity of the null tests to a violation of the equivalence principle is governed by the variance of $\hNe$ and $\hMe$. The measurement of the galaxy number-count fluctuation $\Delta$ is affected by shot noise and by cosmic variance. These uncertainties propagate into the measurement of the multipoles, and hence they affect our null tests. Note that the variance of $\hNe$ and $\hMe$ is due to the variance of each of the multipoles, as well as to the covariance between them. In Appendix~\ref{a:covariance}, we give the explicit expression for the variance of $\hNe$. The variance of $\hMe$ and the covariance between $\hNe$ and $\hMe$ have a very similar form. Here we compute the variances and covariance for a survey with the characteristics of SKA phase 2 (SKA2).

The variance depends on the cosmological parameters, which we set to their best-fit Planck values~\cite{Aghanim:2018eyx}: $\Omega\e{b}=0.04897, \Omega\e{m}=0.3111, n\e{s}=0.9665, A\e{s}=2.105\times 10^{-9}$ and $h=0.6766$. It also depends on the bias of the bright and faint populations, which we model as follows. The mean bias $b(z)$ is given by the specifications of SKA2 (see Table 3 of~\cite{Bull:2015lja}); we then split the galaxies into two populations, with respective biases 
\be
b_\B=b+\Delta b/2, \quad \mbox{and}\quad b_\F=b-\Delta b/2\, .
\ee
The bias difference~$\Delta b$ depends on the population of galaxies targeted by the survey. For BOSS, a bias difference of 1 has been measured between the bright and faint populations of luminous red galaxies~\cite{Gaztanaga:2015jrs}. In the main galaxy sample of SDSS, galaxies have been split into six populations according to their luminosity, with a bias ranging from 0.96 to 2.16~\cite{Percival:2006gt, Cresswell:2008aa}. For the HI galaxies targeted by SKA, the expected bias difference is less well known. In what follows, we will adopt the conservative value $\Delta b=0.5$.

\begin{figure}[!t]
\centering
\includegraphics[width=0.49\columnwidth]{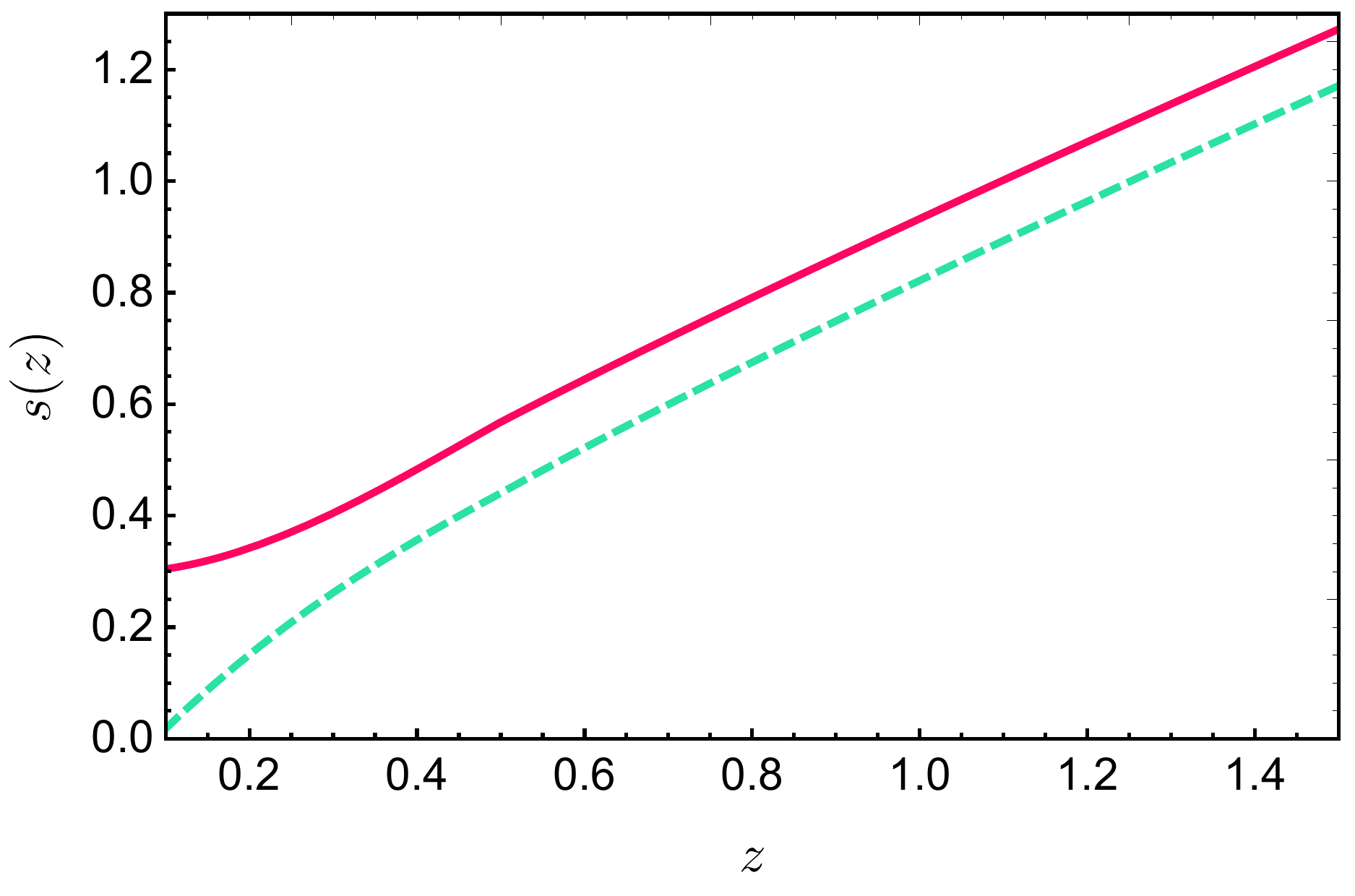}
\caption{Magnification bias of the bright population, $s_\B$ (red solid line), and of the faint population, $s_\F$ (green dashed line), as a function of redshift.}
\label{fig:smag}
\end{figure}

The variance depends also on the values of the magnification bias for the two populations, $s_\B$ and $s_\F$. The magnification bias is given by
\be
s(z,\mF_{\rm lim})=-\frac{2}{5}\frac{\partial\ln \bar{N}(z,\mF>\mF_*)}{\partial\ln \mF_*}\Bigg|_{\mF_*=\mF_{\rm lim}}\, ,
\ee
where $\mF_{\rm lim}$ is the flux limit of the survey, and
\be
\bar{N}(z,\mF>\mF_*)=\int_{\ln \mF_*}^\infty \dd\ln \mF\; \bar{N}(z,\ln \mF) 
\ee
denotes the number density of galaxies with a flux above $\mF_*$. We split the galaxies into two populations with flux above or below $\mF_{\rm cut}$ and we choose $\mF_{\rm cut}$ such that the number density of bright galaxies equals the number density of faint galaxies. In~\cite{Camera:2014bwa}, a fit for $\bar N(z,\mF>\mF_*)$ was provided for SKA2 [see Eq.(A1) and Table~A1 there]. From that fit, we find $\mF_{\rm cut}$ by solving the equation
\be
\bar{N}(z,\mF>\mF_{\rm cut})=\bar{N}(z,\mF>\mF_{\rm lim})-\bar{N}(z,\mF>\mF_{\rm cut})\, .
\ee
The magnification bias of the two populations is then given by
\bea
s_\B(z)=s(z,\mF_{\rm cut})\quad \mbox{and}\quad 
s_\F(z)=s(z,\mF_{\rm lim})\, .
\eea
In Fig.~\ref{fig:smag} we plot the magnification biases as functions of redshift. We see that their difference, $s_\B-s_\F$, ranges from 0.3 at $z=0.15$ to 0.1 at $z=1.5$.

\begin{figure}[!t]
\centering
\includegraphics[width=0.49\columnwidth]{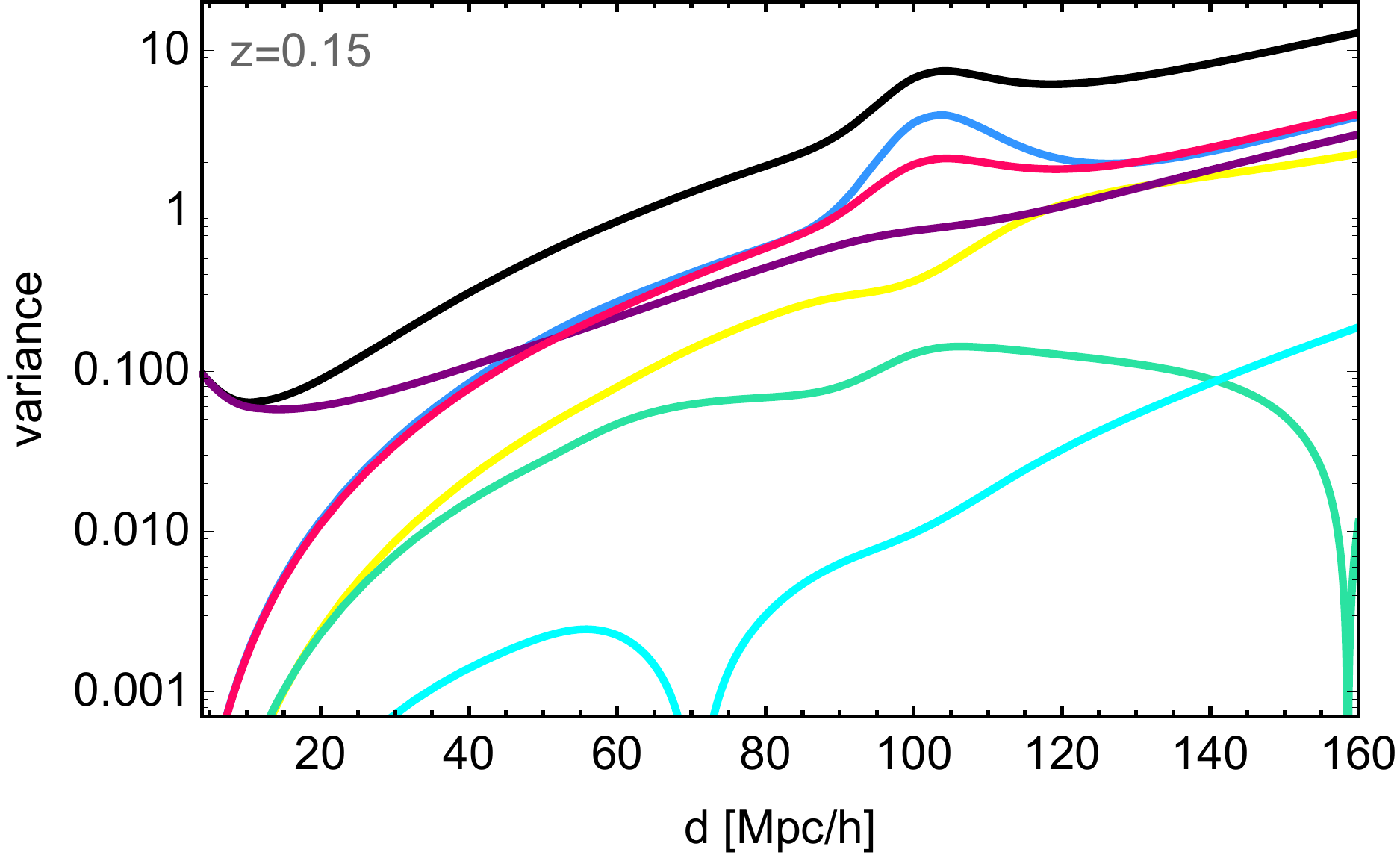}
\includegraphics[width=0.49\columnwidth]{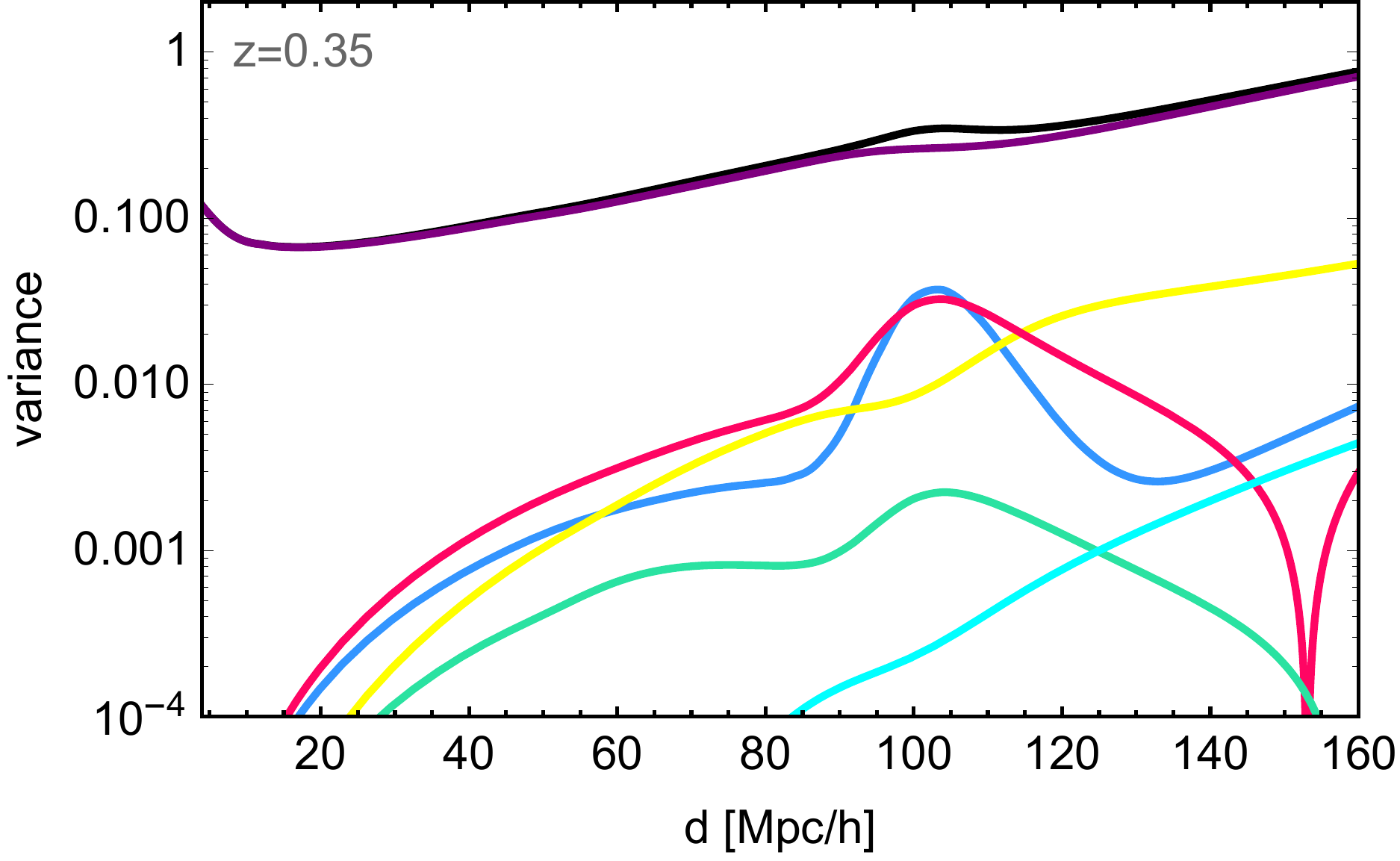}\\
\includegraphics[width=0.49\columnwidth]{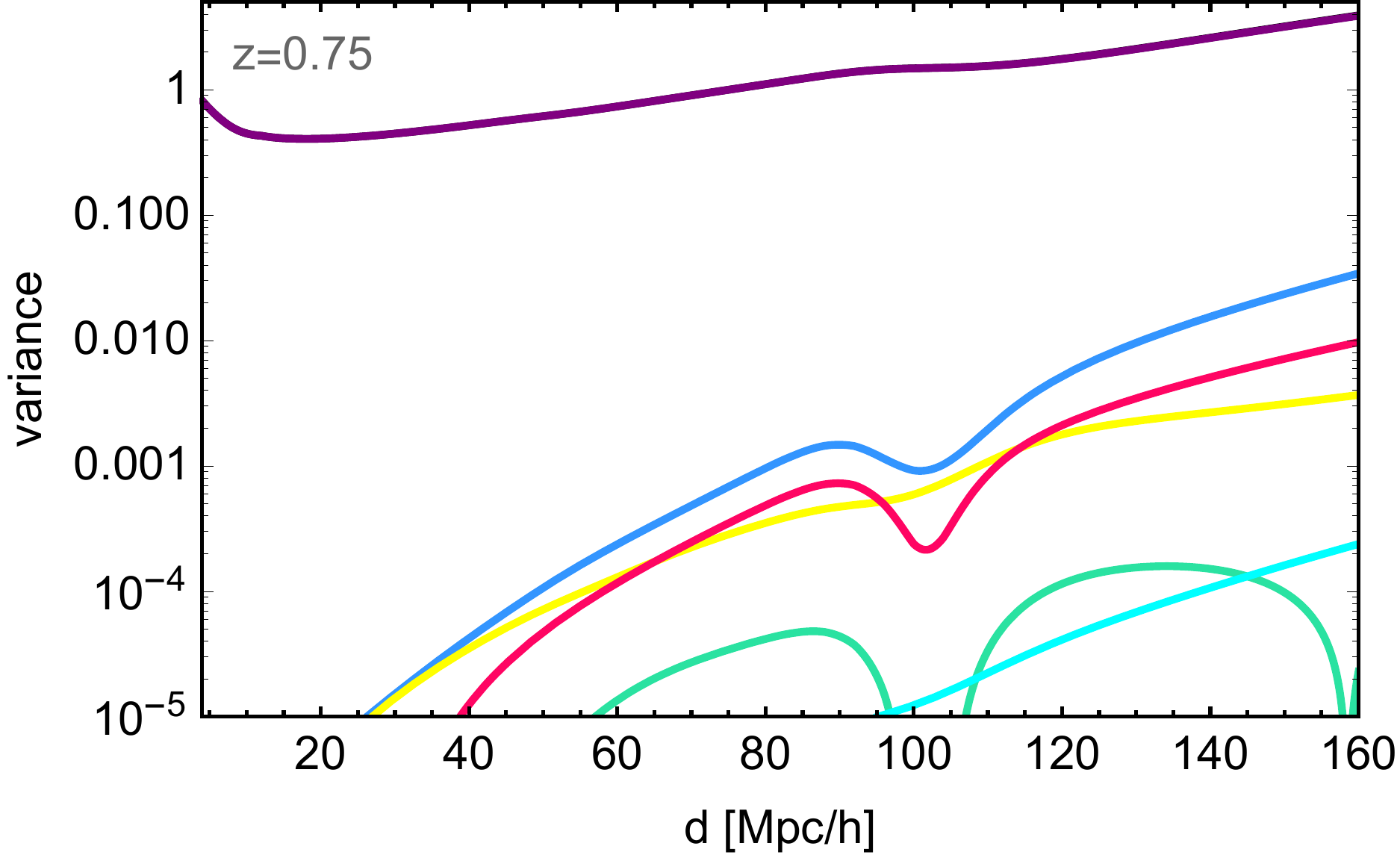}
\includegraphics[width=0.49\columnwidth]{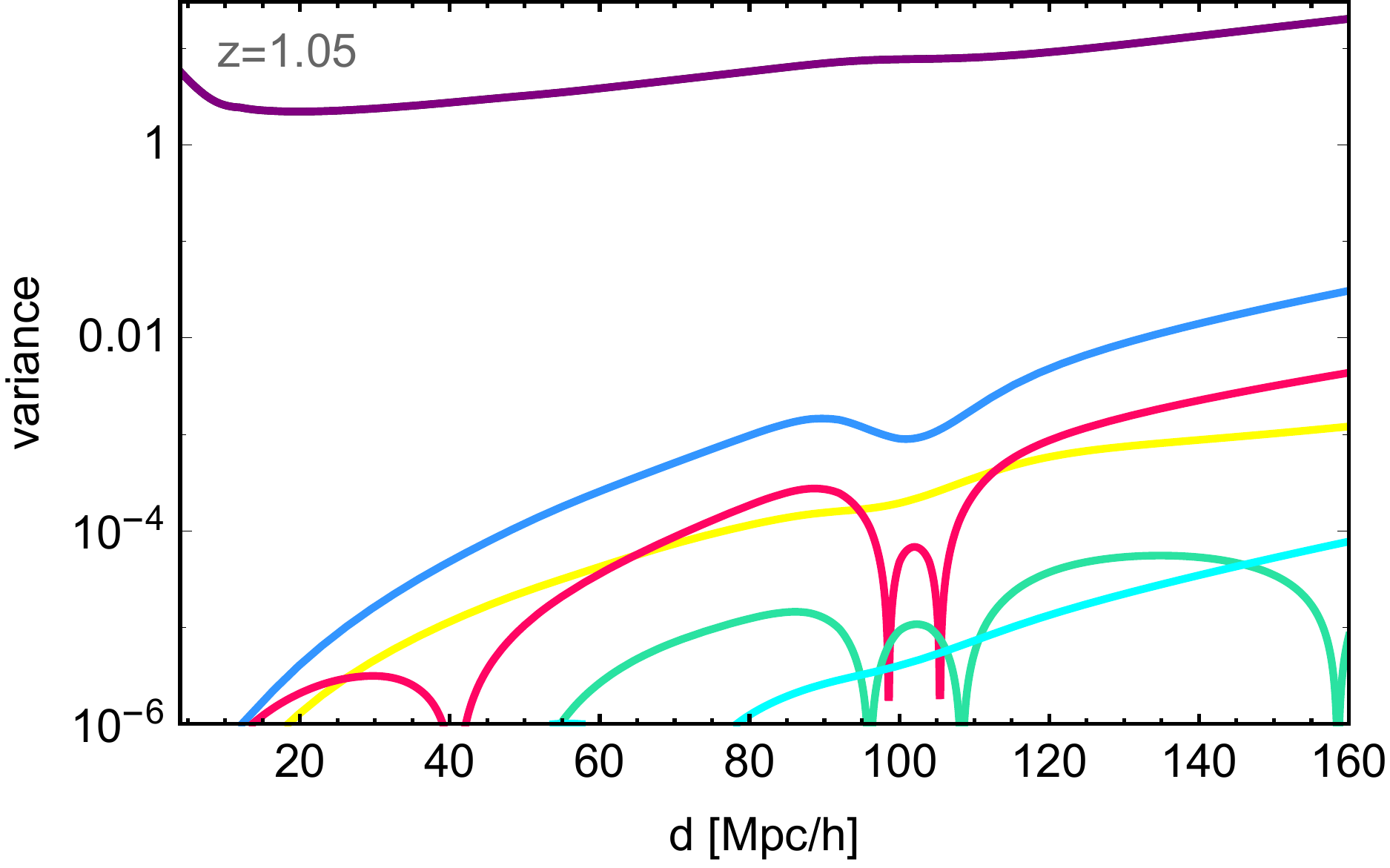}
\caption{Variance of $\hNe$ for various redshift bins, plotted as a function of separation $d$. The black line shows the total, the blue line the quadrupole contribution, the purple line the dipole contribution, and the yellow line the hexadecapole contribution. The covariance between the dipole and the quadrupole is shown in red, the covariance between the dipole and the hexadecapole in cyan and the covariance between the quadrupole and the hexadecapole in green.}
\label{fig:variance}
\end{figure}

Finally, the variance also depends on the evolution biases given by
\be
f^{\rm evol}_\B=\frac{\partial\ln\big( a^3\bar N_\B\big)}{\HH\partial\eta}\quad\mbox{and}\quad
f^{\rm evol}_\F=\frac{\partial\ln\big( a^3\bar N_\F\big)}{\HH\partial\eta}\, ,\label{fevol}
\ee
where $\bar N_\B$ and $\bar N_\F$ denote the mean number density of the bright and faint galaxies respectively. In the following we set these biases to zero. In Sec.~\ref{sec:contaminations}, we will however explore how an uncertainty in these biases can affect the null test.

In Fig.~\ref{fig:variance} we plot the different contributions to the variance of $\hNe$ for a survey like SKA2 for 4 different redshift bins. 
We use a pixel size of 4\,Mpc/$h$. The variance of $\hMe$ is very similar. In the lowest redshift bin, $z=0.15$, we see that the variance is mainly due to the variance of the quadrupole, the variance of the dipole, and the covariance between them. At very large separations the variance of the hexadecapole becomes also important. At larger redshift, $z\geq 0.35$, the variance of the dipole dominates strongly at all separations. 

The variance of the quadrupole is dominated by the cosmic variance of density and RSD. The shot noise contribution to the quadrupole is negligible at all separations and redshifts for a survey like SKA2. On the contrary, due to symmetry reasons, the variance of the dipole is not affected by the pure cosmic variance of density and RSD. The dominant contributions to the variance of the dipole is, therefore, due to the combination of shot noise and cosmic variance.\footnote{Note that in the case where shot noise is negligible, the dominant contribution to the variance of the dipole is due to the cosmic variance of the relativistic effects. We include this contribution here (see Appendix~\ref{a:covariance}), since it is important for the first case studied below, where we neglect shot noise.} As such, it is very sensitive to the number density of galaxies. At low redshift, where the number density of galaxies is high, the cosmic variance of the quadrupole is therefore larger than the mixed shot noise-cosmic variance of the dipole. At larger redshift however, where the number density of galaxies is much smaller and the volume is much larger, the mixed shot noise-cosmic variance of the dipole is the limiting factor. 

\section{Results}
\label{sec:results}

We now forecast the sensitivity of the null test to a violation of the equivalence principle. Even though the null test can be applied on data without any modelling of $\Eb$, at the level of the forecasts, to obtain quantitative results, we need to make some assumptions about $\Eb$. In~\cite{Bonvin:2018ckp}, we have shown that in scalar-tensor and vector-tensor theories of gravity, breaking the equivalence principle results in an additional friction and an extra gravitational-like force acting on dark matter halos. In this case, $\Eb$ takes the form
\be
\Eb=-\HH\Theta(z)\bV\cdot\bn-\Gamma(z)\partial_r\Psi\, ,\label{Ebreak}
\ee
where $\Theta$ encodes the amplitude of the friction term, and $\Gamma$ the amplitude of the extra gravitational-like force. $\Theta$ and $\Gamma$ are dimensionless parameters that directly quantify the amplitude of violation of the equivalence principle. For a specific theory of gravity, these two parameters can be related to the free parameters in the Lagrangian~\cite{Bonvin:2018ckp}. Inserting Eq.~\eqref{Ebreak} into Eq.~\eqref{meanN}, and using the continuity equation to relate $\bV$ and $\delta$, we find that the ensemble average of $\hNe$ and $\hMe$ reads
\begin{align}
\Ne=\Me=(b_\B-b_\F)f\Upsilon\, ,
\end{align}
where 
\be
\Upsilon\define \frac{\Theta-\Gamma}{1+\Gamma} - \frac{\Gamma}{1+\Gamma} \pa{ \frac{\HH'}{\HH^2}+f+\frac{f'}{\HH f} }\, .
\ee

Our goal is to determine how large $\Upsilon$ needs to be to lead to a detectable non-zero $\hNe$ and $\hMe$. Note that for small values of $\Theta$ and $\Gamma$, $\Upsilon$ is linear in these parameters, so it directly encodes the amplitude of violation of the equivalence principle. We construct the Fisher matrix for $\Upsilon$, keeping all other cosmological parameters fixed. We do not vary the cosmological parameters, because our goal is not to measure $\Upsilon$ (which can be degenerate with some of the other cosmological parameters), but rather to determine how large $\Upsilon$ needs to be to reject the null hypothesis $\Ne=\Me=0$. We consider either $\hNe$ alone, or $\hMe$ alone, or the combination of $\hNe$ and $\hMe$. We expect the last case to be more sensitive, because $\hNe$ and $\hMe$ contain different combinations of the quadrupoles~$\hxiq^{\B\B}, \hxiq^{\F\F},\hxiq^{\B\F}$. As such, the reduction of cosmic variance due to combining multiple tracers~\cite{McDonald:2008sh} should apply, and thereby reduce the variance of the quadrupole. 

We first forecast the results of an ideal survey with no shot noise, i.e.\ a cosmic variance limited survey covering $30'000$ square degrees (similar to the sky coverage of SKA2). We assume that the violation of the equivalence principle happens at late times only, such that it does not impact CMB observations, which place stringent constraints on the equivalence of free fall between dark matter and baryons at the time of recombination~\cite{Pettorino:2013oxa}. This assumption is motivated by the intuition that violations of the equivalence principle may be a side effect of the phenomenon driving the acceleration of cosmic expansion. We consider two cases: one where $\Upsilon=\Upsilon_0$ can be considered constant over the redshift range of SKA2 (but somehow decays at higher redshift); and the second where $\Upsilon$ decays towards higher redshift, proportionally to the amount of dark energy
\be
\Upsilon(z)=\Upsilon_0\,\frac{\Omega_\Lambda(z)}{\Omega_{\Lambda}(z=0)}\, ,\label{Upsilonevol}
\ee
where $\Omega_\Lambda(z)$ is the density parameter of the cosmological constant, and $\Upsilon_0$ is constant. 

The Fisher matrix for $\Upsilon_0$, that is a $1\times1$ matrix, reads
\be
\mathcal{F}_{\Upsilon_0}=\sum_{i,j,z}\frac{\partial \hat X(d_i, z)}{\partial \Upsilon_0}\big[{\rm cov}(\hat{X})\big]^{-1}(d_i, d_j, z)\frac{\partial \hat X(d_j, z)}{\partial \Upsilon_0}\, , \nonumber
\ee
where the data vector $\hat X$ is either $\hNe$, $\hMe$ or $(\hNe,\hMe)$, and the sum runs over all the redshift bins of the survey, and over separations $d_i, d_j \in [d_{\rm min},d_{\rm max}]$. We choose $d_{\rm max}=100$\,Mpc/$h$, because for larger distances, we find that cosmic variance gets too large and wipes out all the constraining power of the test. The value of $d_{\rm min}$ is determined by the impact of non-linearities. In Sec.~\ref{sec:nonlinear} we will study how non-linearities affect the null tests using the streaming model to model RSD~\cite{2013MNRAS.431.2834X}. We will see that above $\sim 30$ Mpc/$h$ non-linearities are negligible (compared to the variance of the tests) and linear perturbation theory is valid. We thus choose three representative values for $d_{\rm min}: 20, 32$ and $40$\,Mpc/$h$. We use redshift bins of size $\Delta z=0.1$ from $z_{\rm min}=0.15$ to $z_{\rm max}=1.55$ (here $z$ refers to the mean redshift of the bin). We have checked that beyond $z=1.55$ the signal-to-noise ratio is so small that the constraints do not improve anymore. Note that, in this whole redshift range, the contribution from lensing is completely negligible, as we will show in Section~\ref{sec:lensing}.

The result, $\sigma_{\Upsilon_0}\equiv1/\sqrt{\mathcal{F}_{\Upsilon_0}}$, gives the 1$\sigma$ error bar expected on $\Upsilon_0$. This value is the one that is usually reported when citing the precision with which a parameter can be measured in a future survey. However, the aim of the null test is to differentiate between a theory where the equivalence principle is valid and a theory where the equivalence principle is violated. For this we need to specify with which confidence level we want to perform this test. A 1$\sigma$ difference from zero is not enough to claim an observed violation of the equivalence principle. In Table~\ref{table:forecast_CC}, we report therefore the value of $\Upsilon_0$ that we need to observe a $\hNe$ (or $\hMe$) which is 3$\sigma$ away from zero, i.e.\ we report the value $3\times  \sigma_{\Upsilon_0}$.

\begin{table}[t]
\caption{Value of $\Upsilon_0$ corresponding to a violation of the equivalence principle detectable at $3\sigma$ for a survey which is cosmic variance limited. We show the constraints for a constant $\Upsilon=\Upsilon_0$ and for $\Upsilon$ evolving as in Eq.~\eqref{Upsilonevol}. We use $d_{\rm max}=100$\,Mpc/$h$, $z_{\rm max}=1.55$ and different values for $d_{\rm min}$. \label{table:forecast_CC}}
\begin{center}
\begin{tabular}{| C{1.3cm} |C{1.6cm} C{1.6cm} C{1.8cm} C{1.6cm} C{1.6cm} C{1.6cm} |} 
\hline
 & \multicolumn{3}{c|}{$\Upsilon$=const} &\multicolumn{3}{c|}{$\Upsilon$ evolves}\\
\multicolumn{1}{|c|}{$d_{\rm min}$\, [Mpc/$h$]} & $\Ne$& $\Me$ & \multicolumn{1}{c|}{$\Ne\&\Me$} &$\Ne$& $\Me$ & $\Ne\&\Me$\\ 
\hline 
\multicolumn{1}{|c|}{20}& 0.006 & 0.006 &0.003 & 0.021& 0.020 & 0.011 \\
\multicolumn{1}{|c|}{32}& 0.010 & 0.010 &0.005 & 0.033 & 0.032 & 0.020  \\
\multicolumn{1}{|c|}{40}& 0.012 & 0.012 & 0.006& 0.041 & 0.040 & 0.027 \\
\hline
\end{tabular}
\end{center}
\end{table}

We see that the constraints are about 3 to 4 times tighter in the case where $\Upsilon$ is not evolving. This is not surprising because in this case the violation of the equivalence principle is still relevant at large redshift, where cosmic variance is suppressed due to the larger volume of the redshift bins. The constraints weaken when increasing $d_{\rm min}$. But even for a large $d_{\rm min}=40$\,Mpc/$h$ they are still at the percent level for $\Ne$ and $\Me$ separately. In general the constraints from $\Ne$ and $\Me$ are very similar. This indicates that the sensitivity of the null test to changes in the parameter $\lambda$ in Eq.~\eqref{lambda} may be quite low. Combining $\Ne$ with $\Me$ does however strongly enhances the constraints, by a factor up to 2. This is due to the fact that this combination strongly reduces the cosmic variance of the quadrupole, thanks to the different combinations of bright and faint galaxies used in $\Ne$ and $\Me$.

\begin{table}[t]
\caption{Value of $\Upsilon_0$ corresponding to a violation of the equivalence principle detectable at $3\sigma$ for a survey like SKA2. We show the constraints for a constant $\Upsilon=\Upsilon_0$ and for $\Upsilon$ evolving as in Eq.~\eqref{Upsilonevol}. We use $d_{\rm max}=160$\,Mpc/$h$, $z_{\rm max}=1.05$ and different values for $d_{\rm min}$. \label{table:forecast}}
\begin{center}
\begin{tabular}{| C{1.3cm} |C{1.6cm} C{1.6cm} C{1.8cm} C{1.6cm} C{1.6cm} C{1.6cm} |} 
\hline
 & \multicolumn{3}{c|}{$\Upsilon$=const} &\multicolumn{3}{c|}{$\Upsilon$ evolves}\\
\multicolumn{1}{|c|}{$d_{\rm min}$\, [Mpc/$h$]} & $\Ne$& $\Me$ & \multicolumn{1}{c|}{$\Ne\&\Me$} &$\Ne$& $\Me$ & $\Ne\&\Me$\\ 
\hline 
\multicolumn{1}{|c|}{20}& 0.71 & 0.70 & 0.70 & 1.12& 1.10 & 1.10 \\
\multicolumn{1}{|c|}{32}& 0.83 & 0.82 & 0.82 & 1.33 & 1.31 & 1.31  \\
\multicolumn{1}{|c|}{40}&0.91  & 0.90  & 0.90 & 1.48 & 1.46 & 1.46 \\
\hline
\end{tabular}
\end{center}
\end{table}

We then forecast the results for a survey with the specifications of SKA2. We take the same values as above for the bias, magnification bias and sky coverage. The only difference is that we now include shot noise in the covariance matrix. From Fig.~\ref{fig:variance} we see that in the lowest redshift bin of SKA2, $z=0.15$, the mixed shot noise-cosmic variance contribution from the dipole dominates below 40\,Mpc/$h$. It thus degrades the constraints from small scales. For higher redshift bins however, this contribution strongly dominates at all scales. Therefore, we expect a significant deterioration of the constraints due to shot noise. The results are reported in Table~\ref{table:forecast}. The constraints are weaker by a factor 40-200 compared to the ideal case. The degradation is stronger for smaller $d_{\rm min}$ since cosmic variance is very small at small separations and therefore shot noise has a stronger impact there. We also see that combining $\Ne$ and $\Me$ almost does not improve the constraints. This is due to the fact that when shot noise dominates, reducing the cosmic variance from the quadrupole is not useful. In this regime it is therefore enough to consider only $\Ne$ or $\Me$. To detect a violation of the equivalence principle at 3$\sigma$, we see that we need a parameter $\Upsilon_0$ of the order of 1. Since dark matter has never been detected directly, there is currently no direct constraints on $\Upsilon_0$. As shown in~\cite{Bonvin:2018ckp}, RSD are completely insensitive to this parameter. A value of unity is therefore completely allowed by current data. Note that the constraints on $\Upsilon_0$ are directly proportional to the bias difference between the bright and faint populations, that we have assumed to be relatively small: $b_\B-b_\F=0.5$. If the bias difference turns out to be larger, for example of the order of 1 (as measured in BOSS~\cite{Gaztanaga:2015jrs}), then the constraints would improve by a factor 2.

We can compare our results with the forecasts on a violation of the equivalence principle obtained using consistency relations~\cite{Creminelli:2013nua}. In their forecasts shot noise is neglected, which corresponds to our ideal scenario. In their Fig.\ 1, the 1$\sigma$ constraints for $k_{\rm max}=0.3\,h$/Mpc are of the order of $10^{-4}$, for one redshift bin with a volume of $1\left({\rm Gpc}/h\right)^3$. In our case, the 1$\sigma$ constraint from one redshift bin centered at $z=1.05$ and rescaled to a volume of $1\left({\rm Gpc}/h\right)^3$ is of the order of $ 10^{-2}$, for $d_{\rm min}=2\pi/k_{\rm max}\simeq 20$\,Mpc/$h$. The main difference between the two methods is that our method is sensitive to a change in Euler's equation that would affect all galaxies in a similar way. The two populations of bright and faint galaxies obey the same modified Euler's equation, since we assume that their velocity is driven by the velocity of the dark matter halos. As a consequence, our method is sensitive to the average change in Euler's equation, experienced by the galaxies in the survey. On the contrary, the consistency relation derived in~\cite{Creminelli:2013nua} is sensitive to the difference between the velocity of two populations of galaxies, that would be affected in a different way by a violation of the equivalence principle. In particular, the forecasts presented in Fig.~1 of~\cite{Creminelli:2013nua} rely on the fact that two populations of objects can be selected, one that obeys Euler's equation, and another one that does not. How this can be achieved in practice is not straightforward and will affect the sensitivity of their method. In any case, it would be very interesting to combine the two methods, in order to test the two scenarios. Note that our method would also give a non-zero $\Ne$ and $\Me$ if the two populations of galaxies fall in a different way. This would add an extra contribution to the right-hand side of Eq.~\eqref{meanN}. On the contrary, if all galaxies obey the same modified Euler's equation, the method presented in~\cite{Creminelli:2013nua} would not detect the violation of the equivalence principle. These methods are therefore highly complementary and should be used together. 

\section{Contaminations}
\label{sec:contaminations}

We now explore the limitations of the null test, i.e.\ the situations where $\Ne$ and $\Me$ are different from zero, even if the equivalence principle is valid. 

\subsection{Non-linearities}
\label{sec:nonlinear}

The first source of contamination is due to the impact of non-linear effects on the observables $\hxid, \hxiq^{\B\B}, \hxiq^{\F\F}$ and $\hxih$. To assess the importance of non-linear effects we use the streaming model for RSD that has been used to analyse SDSS data~\cite{2013MNRAS.431.2834X}. This model contains both the Fingers-of-God effect~\cite{Peacock:1993xg} and the impact of non-linearities on the BAO~\cite{Eisenstein:2006nj}. We compute the quadrupole and hexadecapole with this model. For the dipole, we should in principle compute the impact of non-linearities on the gravitational redshift contribution. This has been done using perturbation theory in~\cite{DiDio:2018zmk}. Instead, here, we adopt another approach: we use the linear Euler's equation to relate the gravitational potential to the velocity, and then use the streaming model to compute the impact of non-linearities on the dipole. This procedure is not completely consistent, since non-linear effects are also present in Euler's equation. However, it still gives us an idea of the importance of non-linear effects on the dipole. Since our goal is not to model the dipole precisely, but rather to determine at which scales non-linear effects invalidate the null test, this approach should be sufficient. For the dipole, we also need to compute the impact of non-linearities on the wide-angle corrections. The derivation is shown in Appendix~\ref{a:wide_angle}, using the streaming model. We find that this term is relevant even at small separations. 

\begin{figure}[!t]
\centering
\includegraphics[width=0.49\columnwidth]{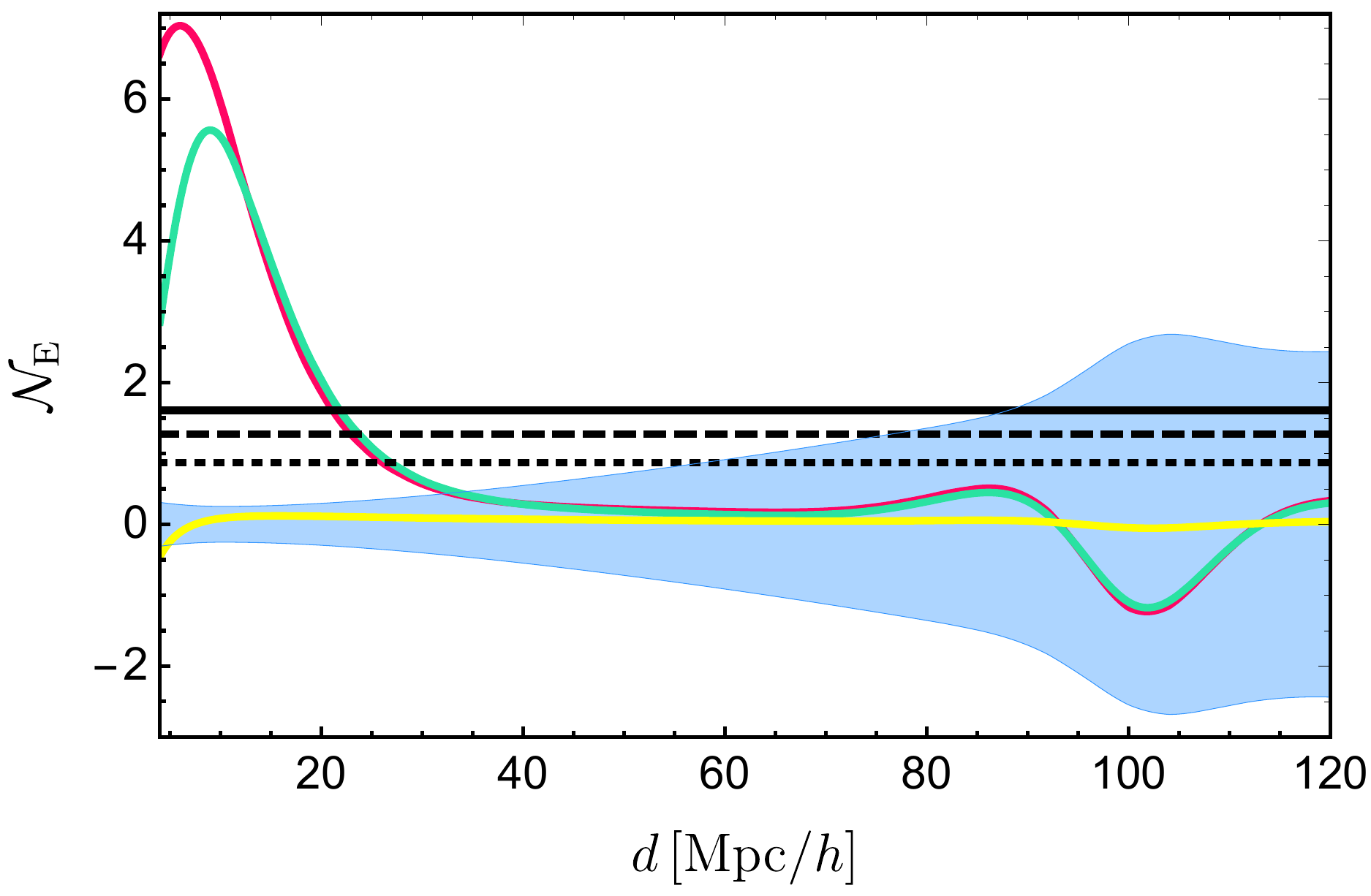}
\includegraphics[width=0.49\columnwidth]{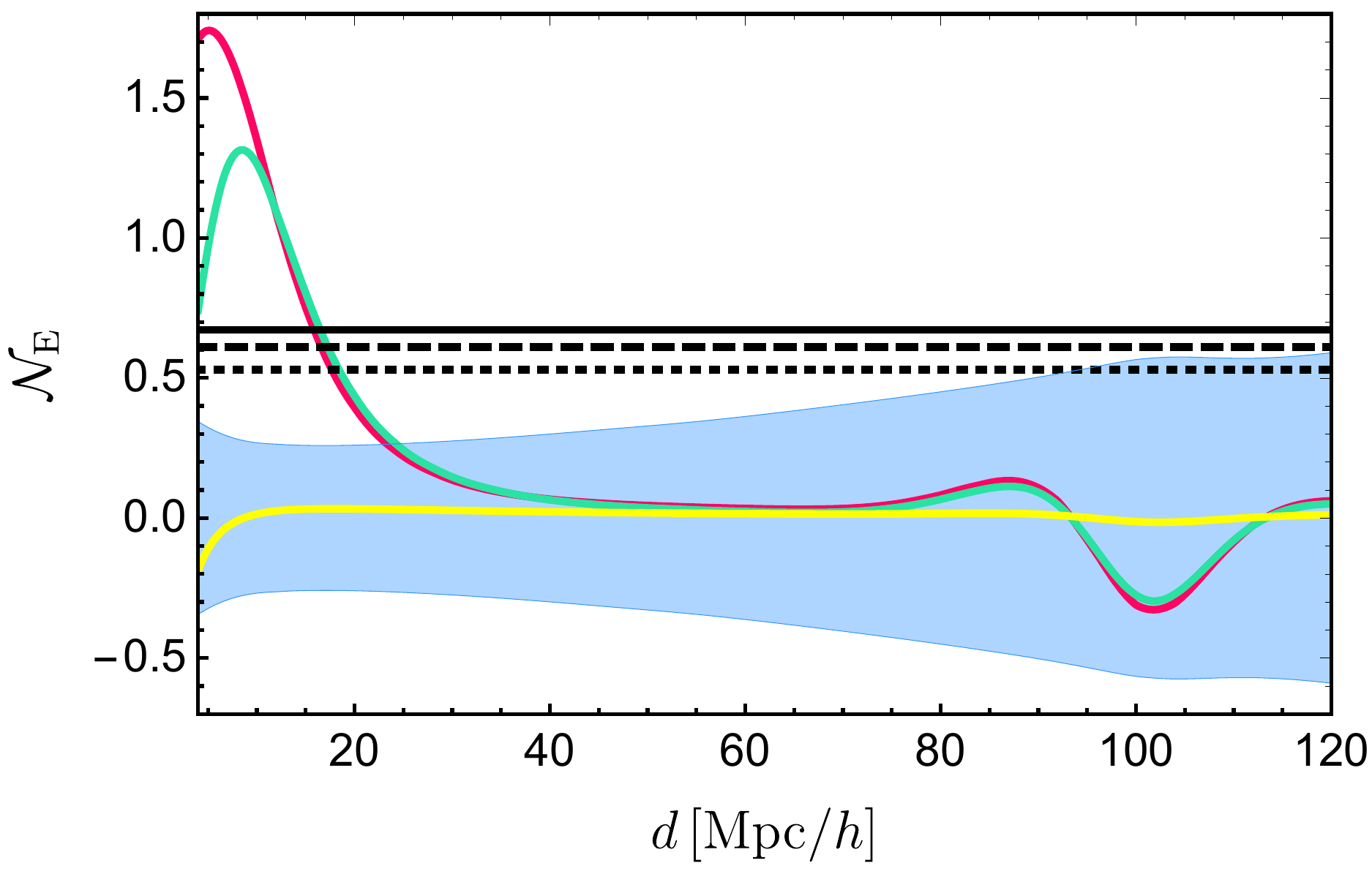}
\caption{Value of the null test, $\Ne$, when the streaming model is used to compute the observables $\hxid, \hxiq^{\B\B}, \hxiq^{\F\F}$ and $\hxih$, and linear perturbation theory is used to compute the coefficients $A, C_\B, C_\F$ and $D$ (red line). For the green line, linear perturbation theory has been replaced by halofit to compute the coefficients. The yellow line shows the null test when non-linear corrections to the bias are included. The black lines show $\Ne$ due to a violation of the equivalence principle, leading to a 3$\sigma$ detection in the redshift bin of consideration, when $d_{\rm min}=20$\,Mpc/$h$ (dotted line), $d_{\rm min}=32$\,Mpc/$h$ (dashed line) and $d_{\rm min}=40$\,Mpc/$h$ (solid line).
The blue shaded region shows the variance of $\hNe$. The left panel is at $z=0.15$ and the right panel at $z=0.35$.}
\label{fig:nonlinear}
\end{figure}

For the coefficients of the null test we use two prescriptions: linear perturbation theory and the halofit power spectrum. The results are shown in Fig.~\ref{fig:nonlinear}, for the lowest redshift bin used in our analysis, $z=0.15$ (left panel), and for $z=0.35$ which is the bin where most of the constraining power comes from (right panel). We see clearly that $\Ne$ does not vanish anymore, due to the presence of non-linear effects. As expected, the impact of non-linearities decreases with redshift. We find that $\Ne$ remains non-zero even if the coefficients are calculated with the halofit power spectrum. This is not surprising, since halofit only accounts for the impact of non-linearities on the density, whereas the null test uses observables that are affected by the peculiar velocity. Because of this, it is not possible to construct a null test that would vanish also in the non-linear regime. The impact of non-linearities on the velocity means that in Eqs.~\eqref{quadBB}--\eqref{hexa} and \eqref{xirel}, the growth rate $f$ cannot be taken out of the integrals over $k$ in $\mu_\ell$ and $\nu_1$. As a consequence the cancellation that takes place in the linear regime when the equivalence principle is valid does not happen anymore in the non-linear regime. 

We compare the value of $\Ne$ using the streaming model, with the variance of the null test. We see that for separations~$d$ larger than $20-40$\,Mpc/$h$ (depending on the redshift), non-linear effects are smaller than the variance. This means that beyond those scales, a detection of a non-zero $\Ne$ cannot be attributed to non-linear effects. For comparison, we also show $\Ne$ due to a violation of the equivalence principle, leading to a 3$\sigma$ detection in the redshift bin of consideration, for different values of $d_{\rm min}$. This clearly indicates that the null test is robust on large scales. 

Another impact of non-linearities is to introduce a scale-dependence in the bias. To assess the impact of this on the null test, we implement the following modelling of the bias in the non-linear regime, used in~\cite{Amendola:2015pha} and fitted from simulations
\be
b(z,k)= b_0(z)\frac{\sqrt{1+Q(z)\left(k/k_1\right)^2}}{\sqrt{1+A(z)k/k_1}}\, ,
\ee
with $A(z)=1.7$, $Q(z)$ fitted from~\cite{Amendola:2015pha}, $k_1=1\,h/$Mpc and $b_0(z)$ from SKA2. This scale-dependence of the bias modifies the multipoles in Eqs.~\eqref{quadBB}, \eqref{quadFF},\eqref{quadBF} and \eqref{xirel} since in this case the bias cannot be taken out of the integrals over $k$ in $\mu_\ell$ and $\nu_1$. As a consequence, $\Ne$ does not vanish, even if the equivalence principle is valid. In Fig.~\ref{fig:nonlinear}, we show the value of $\Ne$ induced by the scale-dependent bias. We see that apart from very small scales ($d\lesssim 5$\,Mpc/$h$) $\Ne$ is always much smaller than the variance. Therefore, the scale-dependence of the bias has no significant impact on the null test.

\subsection{Lensing}

\label{sec:lensing}

\begin{figure}[!t]
\centering
\includegraphics[width=0.49\columnwidth]{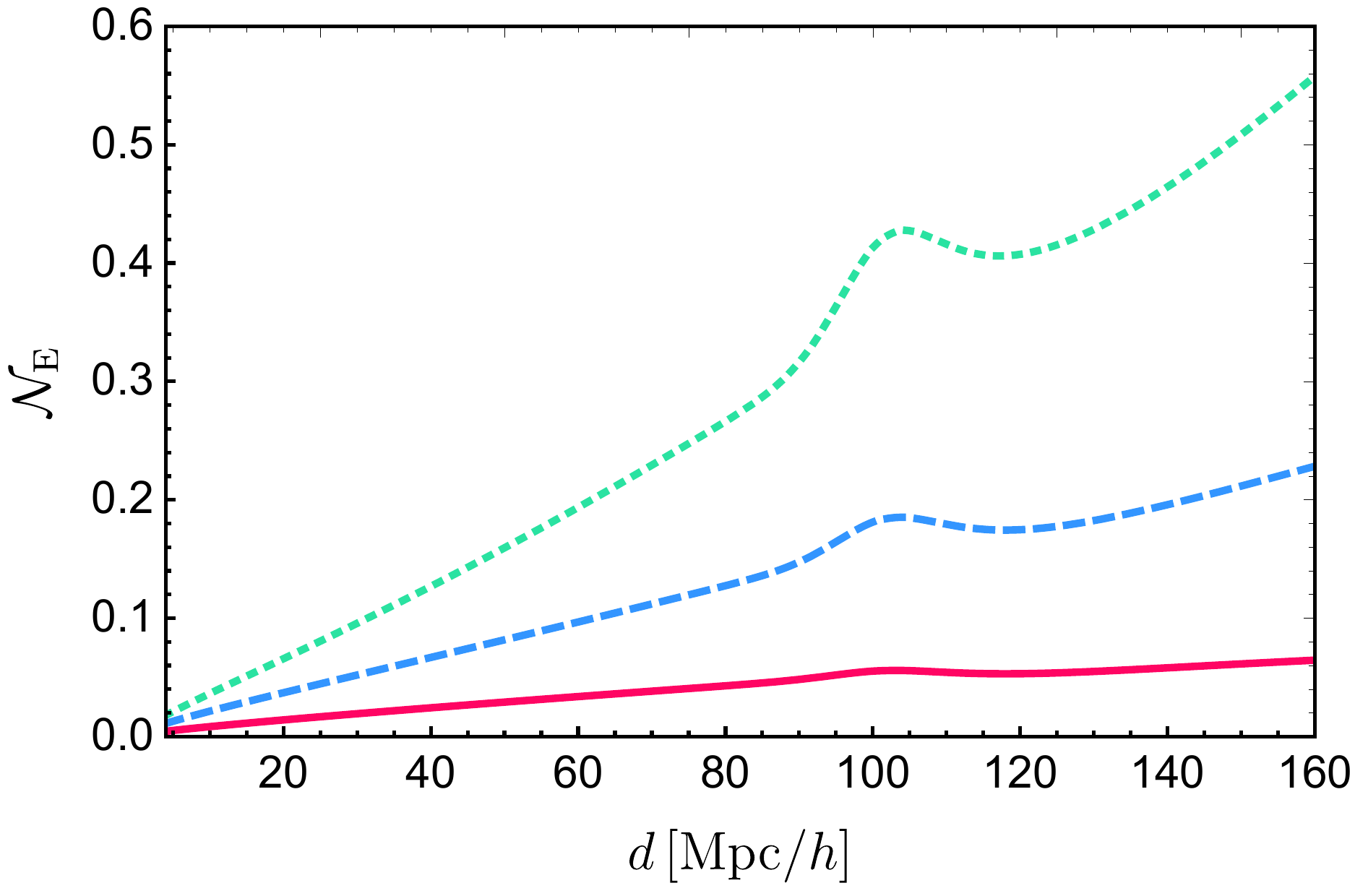}
\includegraphics[width=0.49\columnwidth]{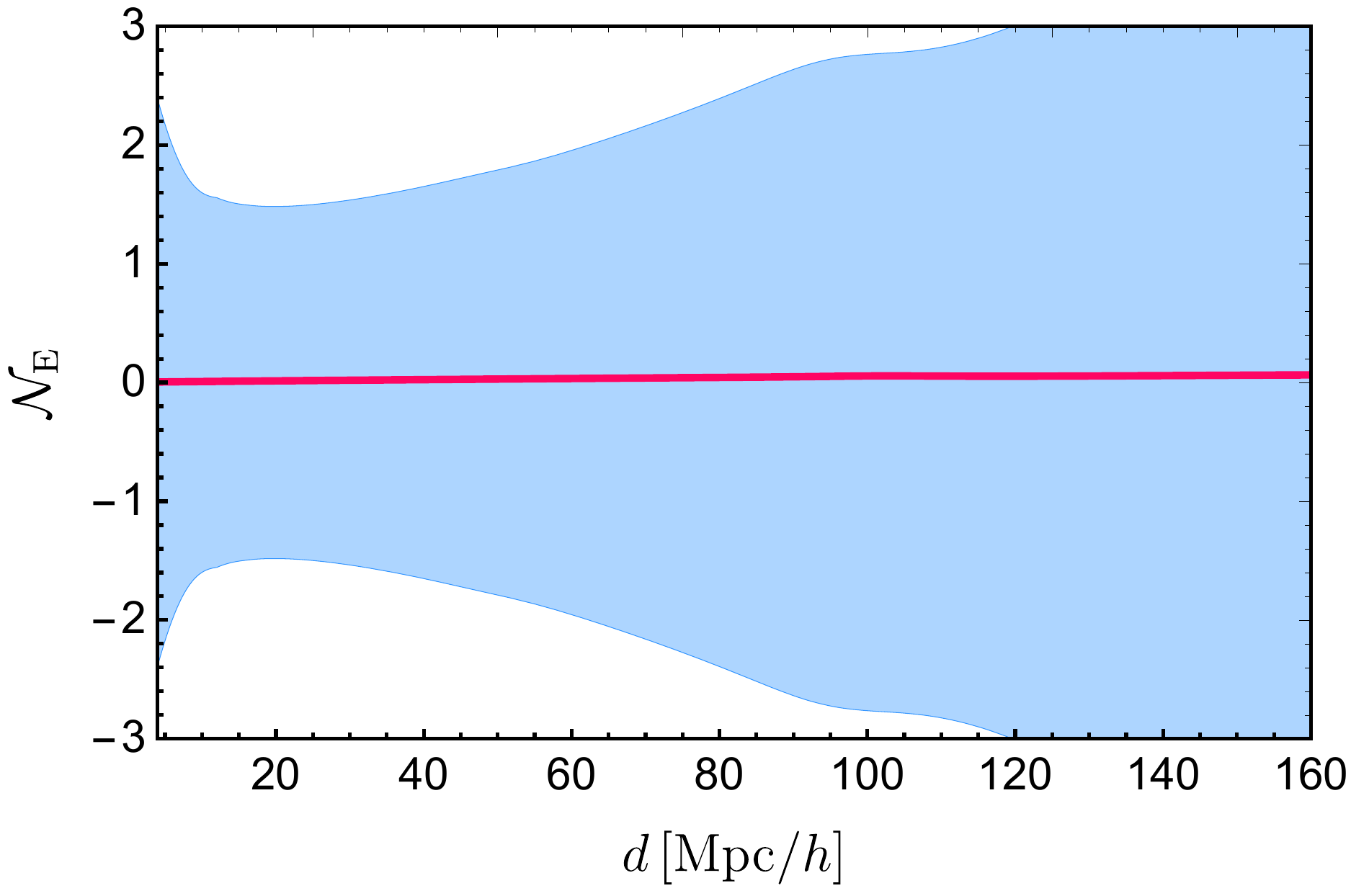}
\caption{{\it Left panel}: lensing contribution to $\Ne$ plotted as a function of separation, at redshift $z=1.05$ (red solid line), $z=1.55$ (blue dashed line), and $z=1.95$ (green dotted line). {\it Right panel}: lensing contribution at $z=1.05$ (red line) plotted with the variance of $\hNe$ (blue shaded region).}
\label{fig:lensing}
\end{figure}

Another source of contamination is lensing magnification, which affects the galaxy number-count fluctuation $\Delta$~\cite{Bonvin:2011bg,Challinor:2011bk}. As such, it contributes to the quadrupole, hexadecapole and dipole and will generate a non-zero $\Ne$ and $\Me$, even when the equivalence principle holds. The contribution from lensing magnification to the correlation function can be calculated using the flat-sky and Limber approximation~\cite{Bonvin:2013ogt}. The multipoles can then be extracted numerically. For two populations of galaxies with luminosity $\L$ and $\L'$, the even multipoles take the form
\begin{align}\label{eq:even_multipoles_xi_lensing}
\xi_\ell^{\L\L'\,{\rm lens}}&= \frac{2\ell+1}{2}
\Bigg\{\frac{3\Omega\e{m}}{4\pi a}\Big[b_\L(5s_{\L'}-2)+b_{\L'}(5s_{\L}-2) \Big]
d\int_0^1 \dd\mu\;\mu P_\ell(\mu)\\
&\times\int \dd k_\perp \; k_\perp \HH_0^2 
P(k_\perp, z)J_0\big(k_\perp d\sqrt{1-\mu^2}\big)\nonumber\\
&+\frac{9\Omega\e{m}^2}{8\pi}(5s_\L-2)(5s_{\L'}-2)\int_0^r \dd r'\;\frac{(r-r')^2r'^2}{r^2 a(r')^2}2\int_0^1\dd\mu\; P_\ell(\mu)\nonumber\\
&\times \int \dd k_\perp \; k_\perp \HH_0^4 
P\big(k_\perp, z(r')\big)J_0\left(k_\perp\frac{r'}{r} d\sqrt{1-\mu^2}\right)\Bigg\}\, ,\quad\mbox{for $\ell$ even}\, ,\nonumber
\end{align}
where $P_\ell$ denotes the order-$\ell$ Legendre polynomial and $J_0$ the order-$0$ Bessel function. The first two lines of Eq.~\eqref{eq:even_multipoles_xi_lensing} contain the density-lensing correlation, which is generated by two contributions. When the galaxy with luminosity $\L'$ is behind the galaxy with luminosity $\L$, the $\L'$ galaxy is lensed by the $\L$ galaxy, and when the galaxy with luminosity $\L$ is behind the galaxy with luminosity $\L'$, the $\L$ galaxy is lensed by the $\L'$ galaxy. The last two lines of Eq.~\eqref{eq:even_multipoles_xi_lensing} contain the lensing-lensing correlation, due to the fact that both galaxies are lensed by the same foreground inhomogeneities. In all the terms, the functions are evaluated at the mean redshift of the bin, and $r$ denotes the conformal distance at that redshift.

The odd multipoles differ from the even ones for two reasons. First, the two density-lensing terms contribute with an opposite sign, which strongly reduces the total contribution. Second, the lensing-lensing contribution exactly vanishes, because it is symmetric in $\mu$. The explicit expression of the odd multipoles is
\begin{align}
\xi_\ell^{\L\L'\,{\rm lens}}&= \frac{2\ell+1}{2}
\Bigg\{\frac{3\Omega\e{m}}{4\pi a}\Big[b_\L(5s_{\L'}-2)-b_{\L'}(5s_{\L}-2) \Big]
d\int_0^1 \dd\mu\;\mu P_\ell(\mu)\\
&\times\int \dd k_\perp \; k_\perp \HH_0^2 
P(k_\perp, z)J_0\big(k_\perp d\sqrt{1-\mu^2}\big)\Bigg\}\, ,\quad\mbox{for $\ell$ odd}\, .\nonumber
\end{align}

In the left panel of Fig.~\ref{fig:lensing}, we show the lensing contribution to $\Ne$ at different redshifts. We see that the contribution quickly increases with redshift, since lensing accumulates along the line-of-sight. In the right panel of Fig.~\ref{fig:lensing}, we compare the lensing contribution at the highest redshift that we are using in the forecasts, $z=1.05$, with the variance of $\hNe$ at that redshift. We see that even though lensing generates a non-zero $\Ne$, this contribution is completely negligible compared to the variance. The null test is therefore unaffected by this effect and can be robustly used at high redshift. Note that even if we go to higher redshift, lensing remains negligible, because the variance due to shot noise increases faster with redshift than the lensing contribution.

\subsection{Scale-dependent growth}

If the growth of structure $f$ does not dependent on $k$, we have seen that $\Ne$ and $\Me$ exactly vanish when the equivalence principle is valid, independently on the form of $f(z)$. This remains true even if gravity is modified and $f$ differs from the $\Lambda$CDM predictions. On the other hand, if $f$ acquires a dependence in $k$, then $\Ne$ and $\Me$ do not vanish anymore. 

In~\cite{Franco:2019wbj}, we proposed another null test, $\mathcal{N}_f$, which combines measurements of the galaxy number-count with measurements of galaxies' sizes, in order to test the scale-independence of $f$ specifically. This null test is completely insensitive to a violation of the equivalence principle, since it involves observables that are only sensitive to the peculiar velocity of galaxies, and not to the gravitational potential $\Psi$. Here we show that a scale-dependence in $f$ that would lead to a 1$\sigma$ deviation from zero in $\mathcal{N}_f$ generates a deviation in $\Ne$ which is slightly below 1$\sigma$. As a consequence, by combining $\mathcal{N}_f$ and $\Ne$ we can unambiguously determine if we have a breaking of the scale-independence of $f$ or a violation of the equivalence principle. 

Following~\cite{Franco:2019wbj}, we parameterise the growth function $D_1(k,z)$ as
\be
D_1(k,z)=\bar D_1(z)\Big[1+\eps(z)\gamma(k) \Big]\, ,\label{Dk}
\ee
where $\bar D_1(z)$ is the growth function in $\Lambda$CDM and
\be
\gamma(k)=c_1\,\frac{1+c_2(k/k_*)^m}{1+(k/k_*)^m}\, ,\quad\mbox{and}\quad \eps(z)=\eps_0\,\frac{\Omega_\Lambda(z)}{\Omega_\Lambda(z=0)}\, .
\ee
We choose to focus on one of the models explored in~\cite{Franco:2019wbj}, with $k_*=0.1 h/$Mpc, $c_1=1, c_2=0$, and $m=-2$. At lowest order in $\eps_0$ we obtain for the growth rate
\be
f(k,z)=\bar f(z)+\gamma(k)\,\frac{\dd\eps(z)}{\dd\ln a}\, ,\label{fk}
\ee
where $\bar f$ denotes the growth rate in $\Lambda$CDM. Inserting~\eqref{Dk} and~\eqref{fk} in the quadrupole, hexadecapole and dipole, we can calculate the resulting $\Ne$.  

In~\cite{Franco:2019wbj}, we found that with SKA2, using $d_{\rm min}=20\,$Mpc/$h$, we can detect at 1$\sigma$ a value of $\eps_0=0.04$ at $z=0.15$. Using this value in $\Ne$, we obtain a non-zero $\Ne$ which is always slightly smaller than the variance, i.e.\ just below detection. At higher redshift, the signal becomes smaller and smaller compared to the variance. This means that an observed deviation from zero in $\Ne$ would automatically lead to a larger observed deviation in $\mathcal{N}_f$. 

\subsection{Wrong fiducial cosmology}

\begin{figure}[!t]
\centering
\includegraphics[width=0.49\columnwidth]{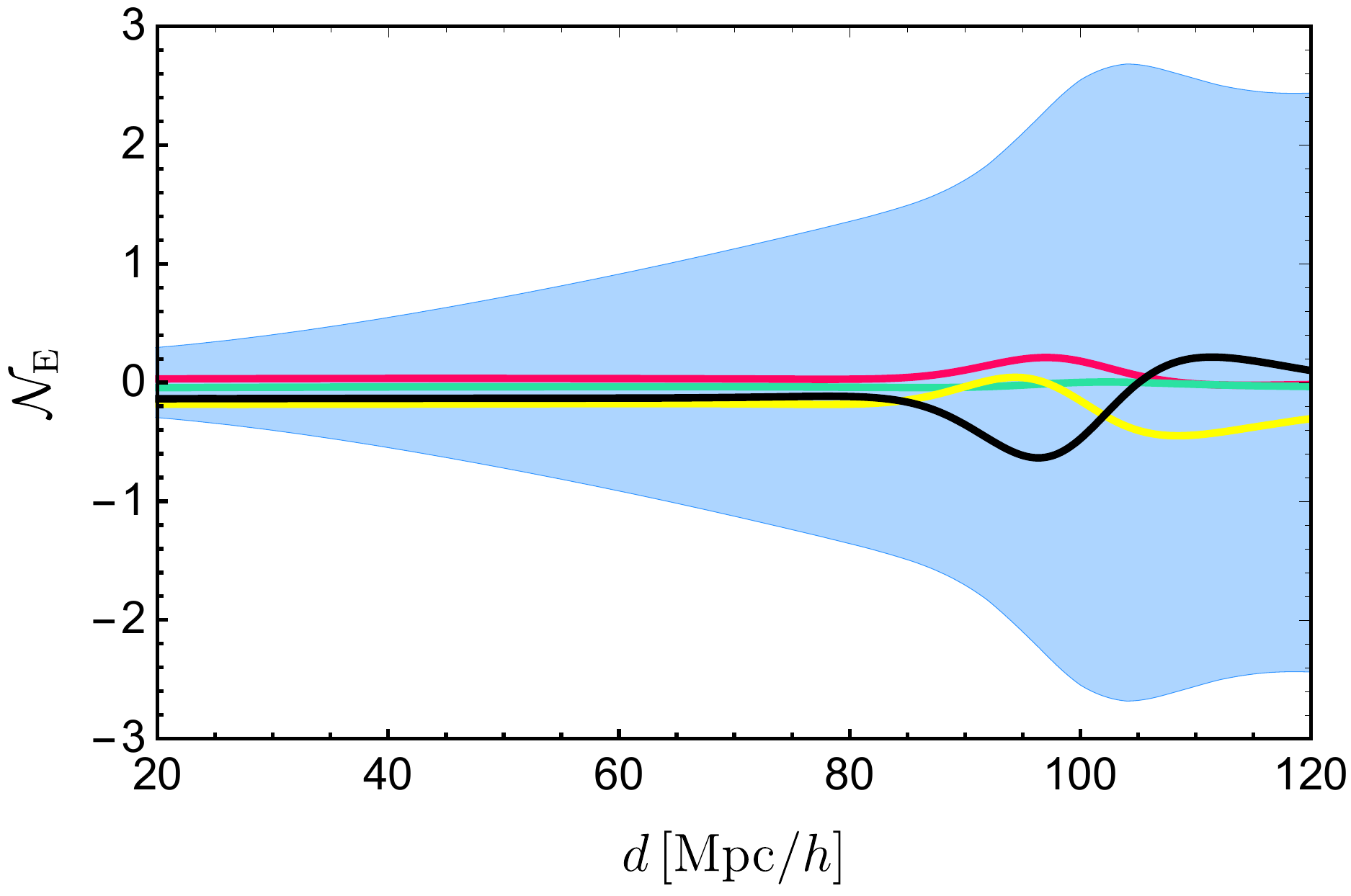}
\includegraphics[width=0.49\columnwidth]{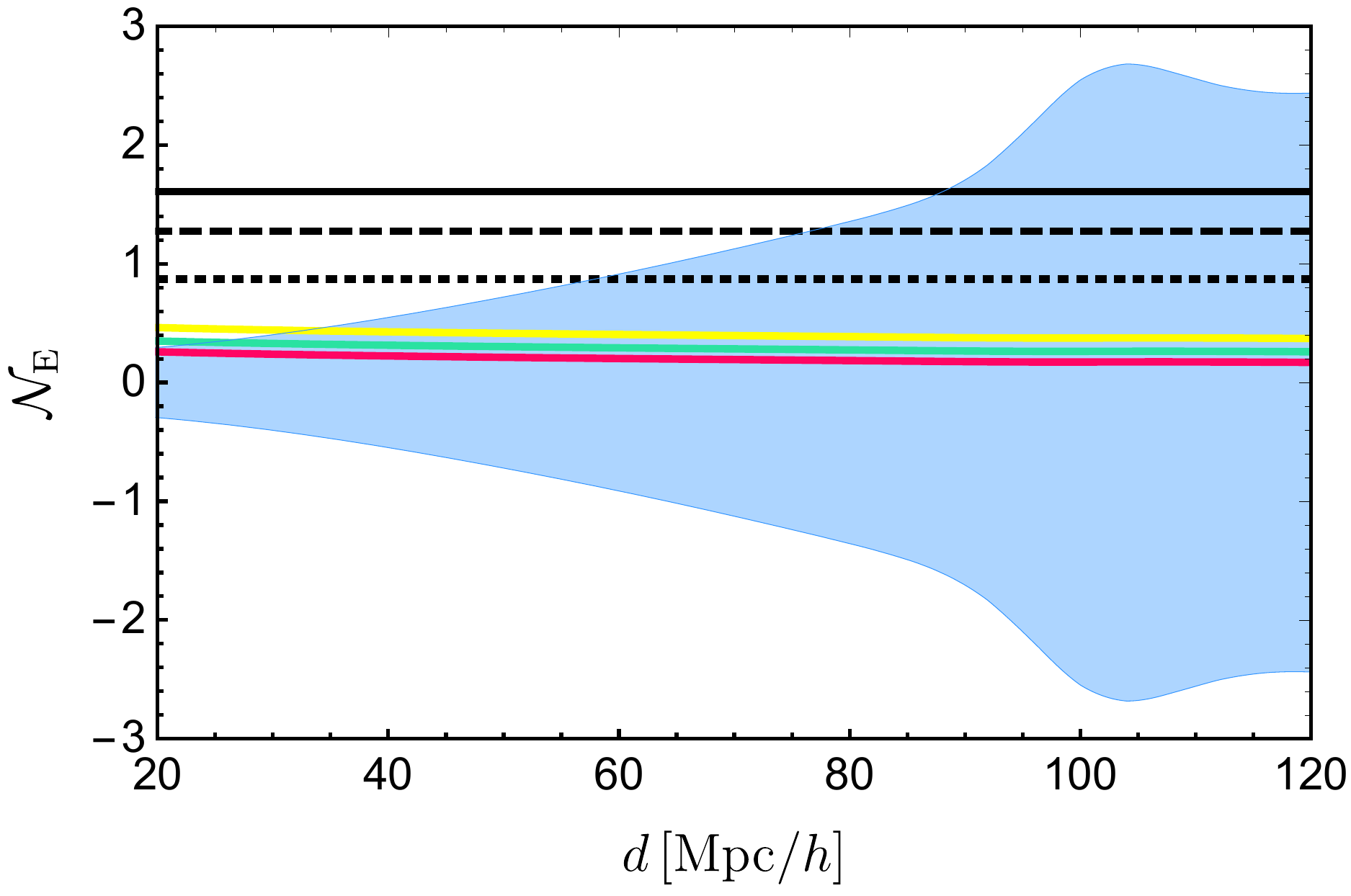}
\caption{{\it Left panel}: Values of $\Ne$ at $z=0.15$ obtained when a wrong fiducial cosmology is used to calculate the coefficients of the test. In red we show the impact of a wrong value of $\Omega\e{b}$, in green of $n\e{s}$, in black of $h$ and in yellow of $\Omega\e{m}$. The blue shaded region shows the variance of $\hNe$. {\it Right panel}: Values of $\Ne$ at $z=0.15$ obtained when a wrong value of the magnification bias $s_\B$ is used to calculate the coefficients of the test. In red we vary $s_\B$ by 1\%, in green by 5\% and in yellow by 10\%. The black lines show $\Ne$ due to a violation of the equivalence principle, leading to a 3$\sigma$ detection at $z=0.15$, when $d_{\rm min}=20$\,Mpc/$h$ (dotted line), $d_{\rm min}=32$\,Mpc/$h$ (dashed line) and $d_{\rm min}=40$\,Mpc/$h$ (solid line). The blue shaded region shows the variance of $\hNe$.}
\label{fig:wrong_cosmo}
\end{figure}

An ideal null test should be independent on cosmological parameters. Here this is however not possible: to construct $\Ne$ we need a certain knowledge about the parameters $\Omega\e{b}, \Omega\e{m}, h$ and $n\e{s}$.\footnote{Note that the primordial amplitude $A\e{s}$ does not affect the null test, since $A\e{s}$ simply factorises out of $\Ne$.} This is indeed necessary to calculate the coefficients of the null test in Eqs.~\eqref{coefftest}, which depend on the comoving distance $\bar r$, the Hubble parameter $\bar \HH$, and the power spectrum $\bar P(k,z)$. If the values of these parameters are wrong, then the null test does not vanish anymore, even if the equivalence principle is valid. However, this turns out not to be a serious issue, because these parameters are already well constrained by Planck. More precisely, we find that even if the parameters $\Omega\e{b}, \Omega\e{m}, h$ and $n\e{s}$ differ from Planck best-fit values by 3$\sigma$, the deviation in $\Ne$ remains much smaller than the variance. Therefore, a wrong choice in the fiducial cosmology does not jeopardise our null test.. 

To calculate this contamination, we compute the coefficients with the fiducial values of $\Omega\e{b}, \Omega\e{m}, h$ and $n\e{s}$, and we compute the observables, $\xi_1^{\B\F}, \xi_2^{\B\B}, \xi_2^{\F\F}$ and $\xi_4$, with the fiducial cosmology + 3$\sigma$ (taken from Table 2 of~\cite{Aghanim:2018eyx}). The null test obtained in this way is shown in the left panel of Fig.~\ref{fig:wrong_cosmo}. We see that the parameter that has the largest impact on the null test is $\Omega\e{m}$. For all parameters, $\Ne$ remains significantly smaller than the variance, and hence any statistically significant observation of a non-zero $\Ne$ cannot be attributed to the wrong choice of a cosmology, except if the real Universe is inconsistent with Planck constraints by significantly more than 3$\sigma$. This result also holds at higher redshift.

\subsection{Uncertainty in the measurement of the magnification bias parameters}

The null test relies on a measurement of the magnification bias parameters $s_\B$ and $s_\F$. These parameters can be inferred by measuring the number density of galaxies in the survey as a function of their flux and redshift. This measurement involves all galaxies, independently on their direction, and hence it is expected to be quite precise. In Fig.~\ref{fig:wrong_cosmo} we show how the null test varies if the value of $s_\B$ that we use to calculate the coefficients of the null test differs from the true value by 1\%, 5\% and 10\% (a similar result holds of course if $s_\F$ varies). We see that above $\sim 30$\,Mpc/$h$, $\Ne$ is smaller than the variance. Below this, the null test is slightly larger than the variance. For comparison, we show also $\Ne$ due to a violation of the equivalence principle, leading to a 3$\sigma$ detection at $z=0.15$ for different values of $d_{\rm min}$. We see that these values are much larger than the ones inferred by a wrong choice of $s_\B$. This result holds also at higher redshift. Therefore, even a large uncertainty of the order of 10\% in the determination of $s$ will not invalidate the null test. 

\subsection{Uncertainty in the evolution biases}

Finally, the coefficients of the null test depend on the evolution biases $\fe_\B$ and $\fe_\F$. From Eq.~\eqref{fevol}, we see that these biases account for the fact that the mean number density of galaxies may evolve with redshift. As such $a^3\bar N_\B$ and $a^3\bar N_\F$ are not constant. These evolution biases can be measured by looking at the evolution of the number densities of bright and faint galaxies as a function of redshift. We have explored how the null test is impacted if the value of $\fe_\B$ (or similarly $\fe_\F$) that we use to calculate the coefficients of the null test differs from the true value. The results depend on the true values. Typically we find that if $\fe_\B$ and $\fe_\F$ are smaller than 2, an uncertainty of $\sim 40\%$ generates a deviation in $\Ne$ that remains smaller than the variance. When $\fe_\B$ and $\fe_\F$ increase, this percentage decreases accordingly, since what matters is the absolute error in the determination of the biases. 

\section{Conclusion}
\label{sec:conclusion}

We proposed a null test to probe the validity of the equivalence principle for dark matter. The null test combines, on the one hand, the quadrupole and hexadecapole of the galaxy correlation function, which are sensitive to the velocity of dark matter halos, and, on the other hand, the dipole of the correlation function, which is sensitive to the time component of the metric $\Psi$, through gravitational redshift. As such, our null test is a direct test of the validity of Euler's equation for dark matter. 

The main advantage of our null test is its model-independence: it does not require any knowledge of the mechanism responsible for the violation of the equivalence principle. Therefore, it can be applied to the data without further assumptions or theoretical modelling. We have demonstrated the robustness of the test against a wide range of possible contaminations, namely non-linear effects, gravitational lensing, and uncertainties in the measurement of cosmological parameters. 

We have forecasted the sensitivity of the null test for the next generation of galaxy surveys. In an ideal survey, where shot noise is negligible and only cosmic variance affects the null test, we have seen that the null test can detect at 3$\sigma$ a violation of the equivalence principle of the order of $10^{-3}$. This constraint degrades significantly if shot noise is included. For a survey with the characteristics of SKA2, a violation of the equivalence principle corresponding to $\Upsilon=0.7$ can be detected at 3$\sigma$, if the bias difference between the two populations of galaxies that are used to measure the dipole is of 0.5. If this bias difference is instead of the order of one, as has been measured in BOSS, then the constraints improve by a factor 2.  In any case, since there are no direct constraints on the validity of the equivalence principle for dark matter and on cosmic scales, the outcome of our test will dramatically improve our knowledge of both gravity and dark matter. 

This null test highlights again the importance of relativistic effects in galaxy clustering as a probe of gravity. The dipole of the correlation function, which is used to build the test, is indeed sensitive to the effect of gravitational redshift, which is not encoded in the so-called Kaiser formula for $\Delta$, but is a key element to probe the validity of the equivalence principle.

\section*{Acknowledgments}
We thank Goran Jelic-Cizmek  for  his  help  using  COFFE.  C.B. and F.O.F. acknowledge  support  by  the  Swiss  National  Science  Foundation. PF received the support of a fellowship from ``la Caixa'' Foundation (ID 100010434). The fellowship code is LCF/BQ/PI19/11690018.

\appendix

\section{Variance}
\label{a:covariance}

We follow the method proposed in~\cite{Hall:2016bmm,Bonvin:2016dze} to calculate the variance of the null tests. The variance contains a shot noise contribution, a cosmic variance contribution and a mixed contribution due to the product of shot noise and cosmic variance. For the cosmic variance we include only the density and RSD terms, since they strongly dominate over the relativistic effects and the lensing contribution. In the terms involving the dipole, the cosmic variance of density and RSD vanishes due to symmetry reasons, but only in the pure cosmic variance contribution, not in the mixed term. In this case, we include the relativistic contributions in the pure cosmic variance. This contribution is strongly subdominant compared to the mixed shot noise-cosmic variance contribution, which contains density and RSD. However, in the ideal case where we neglect shot noise, it is important to include the cosmic variance from relativistic effects, since this is the only contribution. 

We account for the covariance between different separations, $d_i\neq d_j$, but we neglect the covariance between different redshift bins, which is very small since the size of the bins is sufficiently large, $\Delta z=0.1$. The variance depends on the mean number density of the bright and faint galaxies, which we denote respectively by $\bar N_\B$ and $\bar N_\F$, on the volume of the redshift bin, $\VV$, and on the size of the pixels in which $\Delta$ is averaged, which we denote by $\ell\e{p}$. The variance depends also on the following functions
\begin{align}
D_{\ell\ell'}(d_i,d_j)=&\frac{1}{\pi^2}\int \dd k \; k^2 P^2(k,z)j_\ell(kd_i)j_{\ell'}(kd_j)\, ,\\
G_{\ell\ell'}(d_i,d_j)=&\frac{1}{\pi^2}\int \dd k \; k^2 P(k,z)j_\ell(kd_i)j_{\ell'}(kd_j)\, ,\\
H_{\ell\ell'}(d_i,d_j)=&\frac{1}{\pi^2}\int \dd k \; k\HH_0 P^2(k,z)j_\ell(kd_i)j_{\ell'}(kd_j)\, ,\\
I_{\ell\ell'}(d_i,d_j)=&\frac{1}{\pi^2}\int \dd k \; k\HH_0 P(k,z)j_\ell(kd_i)j_{\ell'}(kd_j)\, ,\\
J_{\ell\ell'}(d_i,d_j)=&\frac{1}{\pi^2}\int \dd k \; \HH_0^2 P^2(k,z)j_\ell(kd_i)j_{\ell'}(kd_j)\, ,
\end{align}
and on the coefficient
\be
c_\L=\left(5 s_\L+ \frac{2 - 5 s_\L}{r \HH} +\frac{\HH'}{\HH^2}\right) f\, ,\quad\mbox{for}\quad \L=\B,\F\,.
\ee

Here we show only the results for the variance of $\hNe$. The variance of $\hMe$ and the covariance between $\hNe$ and $\hMe$ are very similar. Below we write separately the different contributions from the dipole, quadrupole and hexadecapole, as well as their covariance. For simplicity we drop the redshift dependence in the expressions. 

\subsection{Quadrupole}

The contribution from the quadrupole contains the variance of $\hxiq^{\B\B}$, of $\hxiq^{\F\F}$ and the covariance between them.

\begin{align}
{\rm cov}_{22}&[\hNe(d_i),\hNe(d_j)]=\left[\frac{C_\B(d_i)C_\B(d_j)}{\bar N_\B^2}+\frac{C_\F(d_i)C_\F(d_j)}{\bar N_\F^2}\right]\frac{5}{2\pi\VV\ell\e{p}d_i^2}\delta_{ij}\label{cov22}\\
&+\frac{2\cdot 5^2}{\VV}\left[\frac{C_\B(d_i)C_\B(d_j)}{\bar N_B}\alpha^{\rm CP}_{22}(b_\B,b_\B)+\frac{C_\F(d_i)C_\F(d_j)}{\bar N_F}\alpha^{\rm CP}_{22}(b_\F,b_\F) \right]G_{22}(d_i,d_j)\nonumber\\
&+\frac{5^2}{\VV}\Big[C_\B(d_i)C_\B(d_j)\alpha^{\rm CC}_{22}(b_\B,b_\B,b_\B,b_\B)+C_\F(d_i)C_\F(d_j)\alpha^{\rm CC}_{22}(b_\F,b_\F,b_\F,b_\F)\nonumber\\
&+\big(C_\B(d_i)C_\F(d_j)+C_\F(d_i)C_\B(d_j)\big)\alpha^{\rm CC}_{22}(b_\B,b_\B,b_\F,b_\F)\Big]D_{22}(d_i,d_j)\, ,\nonumber
\end{align}
where
\begin{align}
&\alpha^{\rm CP}_{22}(b_\L,b_\M)=\frac{1}{5}\left(b_\L b_\M+\frac{11}{21}(b_\L+b_\M)f+\frac{3}{7}f^2 \right) \, ,\\
&\alpha^{\rm CC}_{22}(b_\L,b_\M,b_\N,b_\P)=\frac{1}{5} b_\L b_\M b_\N b_\P + \frac{11}{105} \big[b_\L b_\N b_\P + b_\M b_\N b_\P + b_\L b_\M (b_\N + b_\P)\big] f \\
&+ \frac{3}{35} \big[b_\N b_\P + b_\M (b_\N + b_\P) + b_\L (b_\M + b_\N + b_\P)\big] f^2 + 
 \frac{17}{231} (b_\L + b_\M + b_\N + b_\P) f^3 + \frac{83}{1287} f^4\, .\nonumber
\end{align}
The first line in Eq.~\eqref{cov22} contains the shot noise contribution, the second line contains the mixed shot noise-cosmic variance contribution, and the last two lines contain the cosmic variance contribution.

\subsection{Dipole}

\begin{align}
&{\rm cov}_{11}[\hNe(d_i) ,\hNe(d_j)]=A(d_i)A(d_j)\Bigg\{
\frac{3}{4\pi\bar N_\B\bar N_\F \VV \ell\e{p} d_i^2}\delta_{ij}\label{cov11}\\
&+\frac{9}{2\VV}\left[\frac{1}{\bar N_\F}\left(\frac{b_\B^2}{3}+\frac{2b_\B f}{5}+\frac{f^2}{7} \right)+\frac{1}{\bar N_\B}\left(\frac{b_\F^2}{3}+\frac{2b_\F f}{5}+\frac{f^2}{7} \right)\right]G_{11}(d_i,d_j)\nonumber\\
&+\frac{9}{4\VV}\left(\frac{\HH}{\HH_0}\right)^2\left[\frac{2}{5} (b_\F c_\B - b_\B c_\F)^2 + \frac{4}{7} (c_\B - c_\F) (b_\F c_\B - b_\B c_\F) f + 
 \frac{2}{9} (c_\B - c_\F)^2 f^2\right]J_{11}(d_i,d_j)\Bigg\}\nonumber
\end{align}
The first line contains the shot noise contribution, the second line the mixed shot noise-cosmic variance contribution (where the cosmic variance is due to density and RSD), and the last line contains the cosmic variance contribution, which is due to relativistic effects. Note that the last line is negligible in the presence of shot noise, since it is significantly smaller than the first and second lines. However, in the case where we neglect shot noise, this last line is the only non-zero contribution and we cannot neglect it.

\subsection{Hexadecapole}

\begin{align}
&{\rm cov}_{44}[\hNe(d_i) ,\hNe(d_j)]=D(d_i)D(d_j)\Bigg\{
\frac{9}{2\pi\bar N^2 \VV \ell\e{p} d_i^2}\delta_{ij}\label{cov44}\\
&+\frac{2\cdot9^2}{3\bar N\VV}\left(\frac{b^2}{3}+\frac{26bf}{77}+\frac{643f^2}{5005} \right) G_{44}(d_i,d_j)\nonumber\\
&+\frac{9^2}{\VV}\left(\frac{b^4}{9}+\frac{52b^3f}{231}+\frac{1286b^2f^2}{5005}+\frac{436bf^3}{3003}+\frac{79f^4}{2431} \right)D_{44}(d_i,d_j)\Bigg\}\, .\nonumber
\end{align}
Here $b=(b_\B+b_\F)/2$ is the bias of the whole population of galaxies. The ensemble average of the hexadecapole is independent of bias, but its variance does depend on the value of the bias.
The first line in Eq.~\eqref{cov44} contains the shot noise contribution, the second line the mixed shot noise-cosmic variance contribution, and the last line the cosmic variance contribution.

\subsection{Dipole-quadrupole}

\begin{align}
&{\rm cov}_{12}[\hNe(d_i) ,\hNe(d_j)]=\frac{1}{\VV}\frac{\HH}{\HH_0}
\left[2(b_\F c_\B-b_\B c_\F)+\frac{12}{7}(c_\B-c_\F)f \right]\\
&\times \left[A(d_i)\left(\frac{C_\B(d_j)}{\bar N_\B}+\frac{C_\F(d_j)}{\bar N_\F}\right)I_{12}(d_i,d_j)+d_i\leftrightarrow d_j\right]\nonumber\\
&+\frac{1}{\VV}\frac{\HH}{\HH_0}\Bigg\{2b_\B^2(b_\B c_\F-b_\F c_\B)-\frac{12f}{7}\Big[b_\B^2(c_\B-c_\F)+2b_\B(b_\F c_\B-b_\B c_\F) \Big]\nonumber\\
&-\frac{10f^2}{7}\Big[2b_\B(c_\B-c_\F)+b_\F c_\B-b_\B c_\F \Big]-\frac{40f^3}{33}(c_\B-c_\F)  \Bigg\}
\Big[A(d_i)C_\B(d_j)H_{12}(d_i,d_j)+d_i\leftrightarrow d_j\Big]\nonumber\\
&+\frac{1}{\VV}\frac{\HH}{\HH_0}\Bigg\{2b_\F^2(b_\B c_\F-b_\F c_\B)-\frac{12f}{7}\Big[b_\F^2(c_\B-c_\F)+2b_\F(b_\F c_\B-b_\B c_\F) \Big]\nonumber\\
&-\frac{10f^2}{7}\Big[2b_\F(c_\B-c_\F)+b_\F c_\B-b_\B c_\F \Big]-\frac{40f^3}{33}(c_\B-c_\F)  \Bigg\}
\Big[A(d_i)C_\F(d_j)H_{12}(d_i,d_j)+d_i\leftrightarrow d_j\Big]\nonumber
\end{align}
The first two lines contain the mixed shot noise-cosmic variance contribution, and the last four lines contain the cosmic variance contribution. Note that there is no pure shot noise contribution here, since the dipole contains $\Delta_\B\Delta_\F$ whereas the quadrupole contains $\Delta_\B\Delta_\B$ or $\Delta_\F\Delta_\F$. Shot noise only affects correlations of $\Delta$ for the same population of galaxies, and therefore the pure shot noise term exists only if we have an even number of $\Delta_\B$ and an even number of $\Delta_\F$, which is not the case here.

\subsection{Dipole-hexadecapole}

The covariance between the dipole and the hexadecapole is significantly smaller than the other contributions. Moreover, since the number density of galaxies is large in SKA2, the shot noise contribution and the mixed contribution are always negligible compared with the cosmic variance contribution (apart in the dipole term, because there the cosmic variance contribution of density and RSD vanishes). Here we neglect therefore the shot noise and mixed term and show only the pure cosmic variance contribution
\begin{align}
&{\rm cov}_{14}[\hNe(d_i) ,\hNe(d_j)]=\frac{27}{2\VV}\frac{\HH}{\HH_0}
\Big[\frac{16}{315}  \Big(2 b_\F c_\B + b (c_\B - c_\F) - 2 b_\B c_\F\Big) bf \\
&+ \frac{16}{231} \Big(b_\F c_\B + 2 b (c_\B - c_\F) - b_\B c_\F\Big) f^2 + \frac{32}{429} (c_\B - c_\F) f^3\Big]
\Big[A(d_i)D(d_j)H_{14}(d_i,d_j)+d_i\leftrightarrow d_j\Big]\, .\nonumber
\end{align}

\subsection{Quadrupole-hexadecapole}

The covariance between the quadrupole and the hexadecapole is also significantly smaller than the other contributions. Therefore, as above, we only show the dominant contribution due to cosmic variance
\begin{align}
&{\rm cov}_{24}[\hNe(d_i) ,\hNe(d_j)]=-\frac{45}{2\VV}D_{24}(d_i,d_j)\\
&\times\Bigg\{
\Bigg[\frac{16f}{105} b\, b_\B (b + b_\B)  + \frac{272f^2}{3465} (b^2 + 4 b\,b_\B + b_\B^2) 
+\frac{464f^3}{3003} (b + b_\B)  + \frac{32 f^4}{429}\Bigg]C_\B(d_i)D(d_j)\nonumber\\
&+\Bigg[\frac{16f}{105} b\, b_\F (b + b_\F)  + \frac{272f^2}{3465} (b^2 + 4 b\, b_\F + b_\F^2)
+\frac{464f^3}{3003} (b + b_\F)  + \frac{32 f^4}{429}\Bigg]C_\F(d_i)D(d_j)\Bigg\}\nonumber\\
&+d_i\leftrightarrow d_j\, .\nonumber
\end{align}

\section{Wide-angle effects in the non-linear regime}
\label{a:wide_angle}

\begin{figure}[!h]
\centering
\includegraphics[width=0.49\columnwidth]{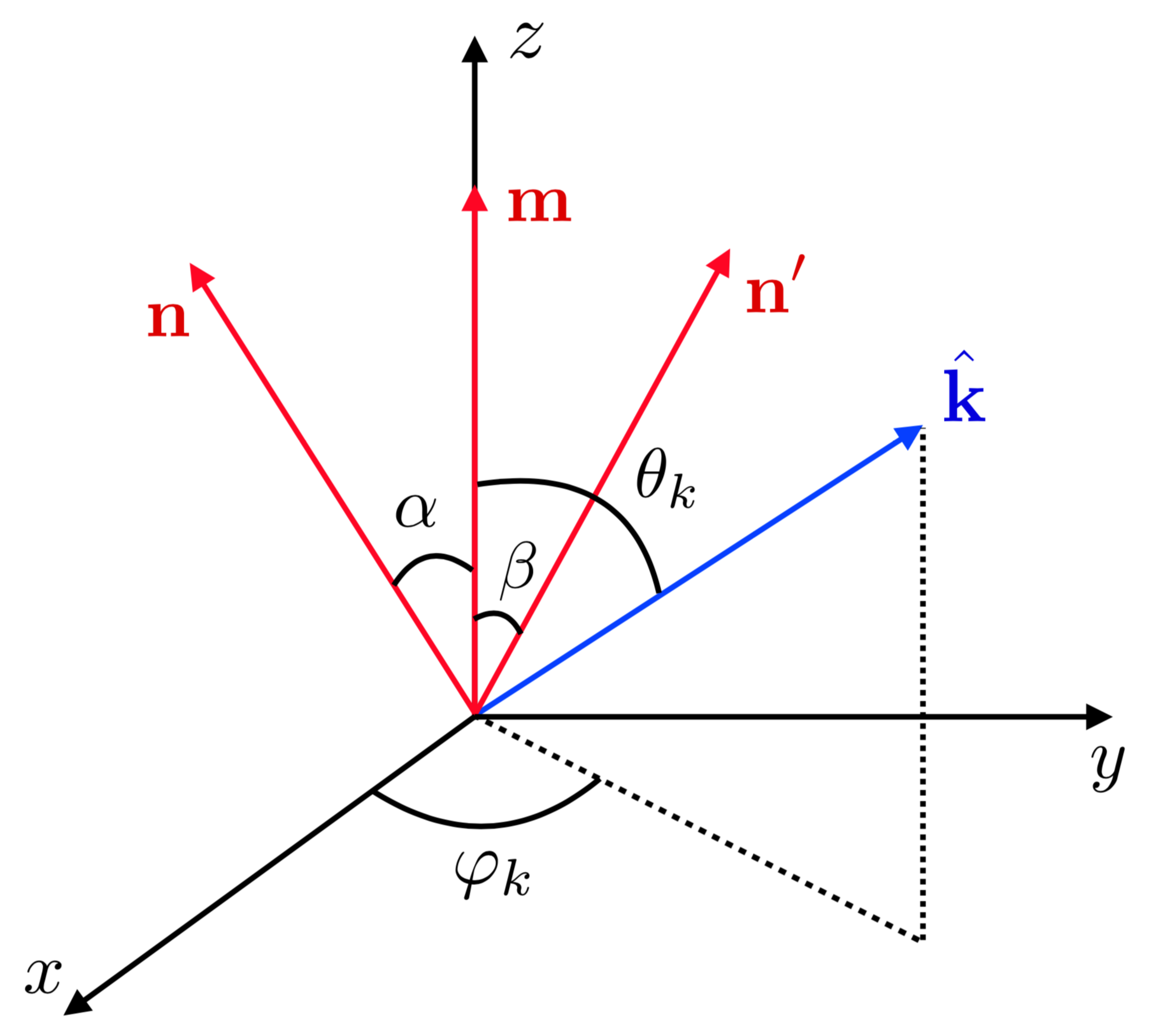}\hspace{0.5cm}
\includegraphics[width=0.37\columnwidth]{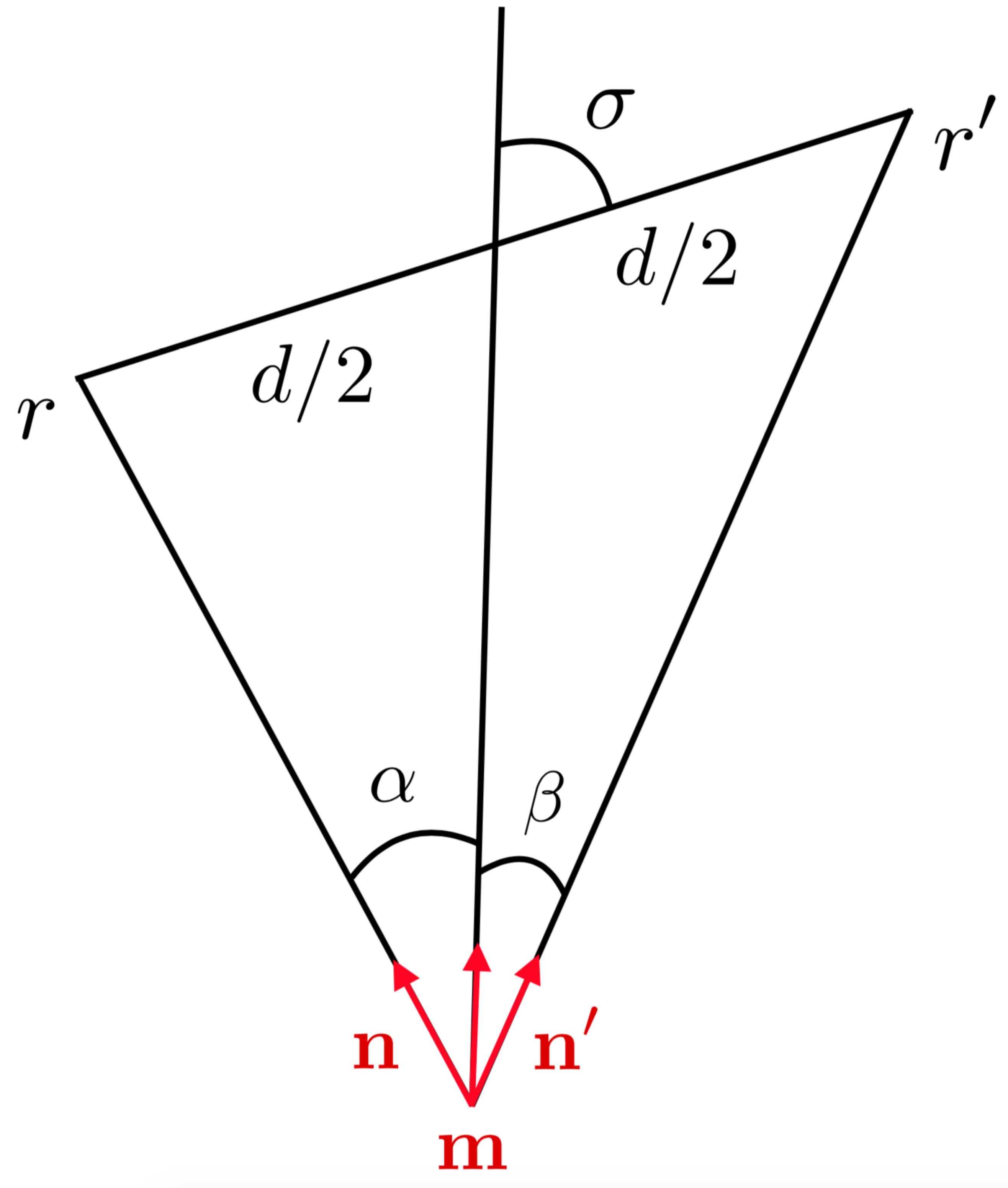}
\caption{Representation of the different angles and distances relevant for the calculation of the wide-angle effects.}
\label{fig:wide}
\end{figure}

We derive here the wide-angle contribution to the dipole, using the streaming model presented in~\cite{2013MNRAS.431.2834X}. The wide-angle contribution is due to the fact that in the full sky, RSD depend on the two line-of-sights $\bn$ and $\bn'$ to the two galaxies in the pair, which are not exactly the same.  The cross-correlation between bright and faint galaxies can be written as
\bea
\langle\Delta_\B^{\rm st}(z,\bn)\Delta_\F^{\rm st}(z',\bn') \rangle&=&\frac{1}{(2\pi)^3}\int \dd^3\bk \; \ex{\i\bk(\bx'-\bx)}\Big(b_\B(z)+f(z)(\hat{\bk}\cdot\bn)^2 \Big)\Big(b_\F(z')+f(z')(\hat{\bk}\cdot\bn')^2 \Big)\nonumber\\
&&\times P^{\rm str}(k,\hat{\bk}\cdot \bmm,z)\, ,
\eea
where $\bmm$ denotes the direction to the middle point between the galaxies (see Fig.~\ref{fig:wide}) and $\Pstr$ is the power spectrum in the streaming model given by~\cite{2013MNRAS.431.2834X}
\be
\Pstr(k,\hat{\bk}\cdot \bmm,z)=F(k,\hat{\bk}\cdot \bmm, \Sigma\e{s})\left[\frac{D_1(z)}{D_1(z=0)} \right]^2P_{\rm dw}(k,\hat{\bk}\cdot \bmm,z=0)\, .
\ee
Here $F$ describes the Fingers-of-God effect
\be
F(k,\hat{\bk}\cdot \bmm, \Sigma\e{s})=\frac{1}{\Big[1+k^2(\hat{\bk}\cdot \bmm)^2\Sigma\e{s}^2\Big]^2}\, ,
\ee
with $\Sigma\e{s}$ the streaming scale. $P_{\rm dw}$ is the de-wiggled power spectrum defined in~\cite{Eisenstein:2006nj}, which depends on two parameters $\Sigma_\parallel$ and $\Sigma_\perp$. This term encompasses the effect of non-linear effects on the BAO. For our calculation we choose the values $\Sigma\e{s}=4$\,Mpc/$h$, $\Sigma_\parallel=10$\,Mpc/$h$ and $\Sigma_\perp=6$\,Mpc/$h$, used in~\cite{2013MNRAS.431.2834X}.

In the following we neglect the evolution corrections, due to the fact that the bias and the growth function evolve with redshift and we set $z'=z$. These terms have been shown to be much smaller than the relativistic dipole~\cite{Bonvin:2013ogt}. We account however for the fact that $\bn\neq\bn'\neq\bmm$.

The dipole contribution from the wide-angle effects is given by
\begin{align}
\label{dipwide_gen}
&\xi_1^{\B\F\, {\rm wide}}=\frac{3}{2}\int_{-1}^1 \dd\mu \; P_1(\mu)\frac{1}{2}\Big[\langle\Delta_\B^{\rm st}(z,\bn)\Delta_\F^{\rm st}(z,\bn') \rangle-\langle\Delta_\B^{\rm st}(z,\bn')\Delta_\F^{\rm st}(z,\bn) \rangle \Big] \\
&=\frac{3}{4(2\pi)^3}(b_\B-b_\F)f\int_{-1}^1 \dd\mu \; P_1(\mu)\int \dd^3\bk \; \ex{\i\bk(\bx'-\bx)}
\Big[(\hbk\cdot\bn')^2-(\hbk\cdot\bn)^2 \Big]\Pstr(k,\hat{\bk}\cdot \bmm,z)\,,\nonumber
\end{align}
with $\mu=\cos\sigma$. In the flat-sky approximation we have $\hbk\cdot\bn=\hbk\cdot\bn'=\hbk\cdot\bmm$ and Eq.~\eqref{dipwide_gen} vanishes. Here we want to calculate the lowest order correction to the flat-sky approximation, i.e.\ we express $\hbk\cdot\bn$ and $\hbk\cdot\bn'$ in terms of $\hbk\cdot\bmm$ with corrections of the order $d/r$. The vectors $\bn, \bn'$ and $\bmm$ are in the same plane, that we choose to be the $(yz)$-plane, as shown on Fig.~\ref{fig:wide} (left panel). Without loss of generality, we choose $\bmm$ on the $z$ axis, and we denote by $\alpha$ and $\beta$ the angles between $\bmm$ and $\bn$, and $\bmm$ and $\bn'$ respectively. Note that since $\bmm$ is the vector pointing in the direction of the middle point between the two galaxies, $\alpha$ is not exactly equal to $\beta$. We find
\begin{align}
\hbk\cdot\bn&=-\sin\theta_k\sin\varphi_k\sin\alpha+\cos\theta_k\cos\alpha\, ,\\
\hbk\cdot\bn'&=\sin\theta_k\sin\varphi_k\sin\beta+\cos\theta_k\cos\beta\, ,
\end{align}
where $(\theta_k,\varphi_k)$ are the angles defining the direction of $\bk$.

We can then express $\alpha$ and $\beta$ in terms of the angle $\sigma$ which denotes the orientation of the pair of galaxies with respect to $\bmm$, see Fig.~\ref{fig:wide} (right panel). We find
\begin{align}
&\sin\alpha=\sin\beta=\frac{d}{2r}\sin\sigma+\mathcal{O}\left(\frac{d}{r} \right)^2\quad\mbox{and}\quad \cos\alpha=\cos\beta=1+\mathcal{O}\left(\frac{d}{r} \right)^2 \, . \label{alphabeta}
\end{align}

We insert Eq.~\eqref{alphabeta} into Eq.~\eqref{dipwide_gen} and expand
\begin{align}
\ex{\i\bk(\bx'-\bx)}&=\ex{\i\bk\mathbf{N}kd}
=4\pi\sum_{\ell m}\i^\ell Y_{\ell m}(\bN)Y^*_{\ell m}(\theta_k,\varphi_k)j_\ell(kd)\\
&=4\pi\sum_{\ell m}\i^\ell\sqrt{\frac{(2\ell+1)(\ell-m)!}{4\pi(\ell+m)!}}P^m_\ell(\cos\theta_k)\ex{-\i m\varphi_k} Y_{\ell m}(\bN)j_\ell(kd)\, ,\nonumber
\end{align}
where $\bN$ is a unit vector along the pair of galaxies with spherical coordinates $(\sigma,\pi/2)$, and $P^m_\ell$ are the associated Legendre polynomials. The integrals over $\varphi_k$ and $\sigma$ can be done analytically. We have
\be
\int_0^{2\pi}\dd\varphi_k \; \sin\varphi_k \ex{-\i m\varphi_k}=-\i\pi\delta_{m1}+\i\pi\delta_{m-1}\, ,
\ee
and
\be
\int_0^\pi \dd\sigma\;\sin^2\sigma P_1(\cos\sigma)Y_{\ell\pm1}(\sigma,\pi/2)=-\i\sqrt{\frac{2}{15\pi}}\delta_{\ell 2}\, .
\ee
Using that $P^{-1}_2(\cos\theta_k)=-\frac{1}{6}P^{1}_2(\cos\theta_k)=-3\cos\theta_k\sin{\theta_k}$, we find
\begin{align}
\xi_1^{\B\F\, {\rm wide}}=&-\frac{3}{4\pi^2}(b_\B-b_\F)f\left[\frac{D_1(z)}{D_1(z=0)} \right]^2\frac{d}{r}\int_{-1}^1\dd\mu\; \mu^2(1-\mu^2)\\
&\times \int \dd k \; k^2 j_2(kd)F(k,\mu,\Sigma_s)P_{\rm dw} (k,\mu, z=0)\, .\nonumber
\end{align}

\bibliographystyle{JHEP.bst}
\bibliography{bibliography_Euler.bib}

\end{document}